\documentclass[a4paper,onecolumn,11pt,accepted=2025-11-24]{quantumarticle}
\pdfoutput=1
\usepackage[utf8]{inputenc}
\usepackage[english]{babel}
\usepackage[T1]{fontenc}
\usepackage{amsmath}
\usepackage{hyperref}

\usepackage{listings}

\PassOptionsToPackage{compress}{natbib}
\usepackage[numbers]{natbib}

\usepackage{braket}
\usepackage{amsfonts}
\usepackage{svg}
\usepackage{subfig}

\usepackage{tikz}
\usepackage{lipsum}

\usepackage{cleveref}

\begin{document}

\title{Programming tools for Analogue Quantum Computing in the High-Performance Computing Context -- A Review}

\author[1,2]{Mateusz Meller}
\author[1]{Vendel Szeremi}
\author[2]{Oliver Thomson Brown}

\affil[1]{Hartree Centre, Science and Technology Facilities Council (STFC), UK Research and Innovation (UKRI)}
\affil[2]{EPCC, The University of Edinburgh, UK}
\orcid{0000-0002-5193-8635}

\maketitle
\begin{abstract}
Recent advances in quantum computing have brought us closer to realizing the potential of this transformative technology. While significant strides have been made in quantum error correction, many challenges persist, particularly in the realm of noise and scalability. Analogue quantum computing schemes, such as Analogue Hamiltonian Simulation and Quantum Annealing, offer a promising approach to address these limitations. By operating at a higher level of abstraction, these schemes can simplify the development of large-scale quantum algorithms.
To fully harness the power of quantum computers, they must be seamlessly integrated with traditional high-performance computing (HPC) systems. While substantial research has focused on the integration of circuit-based quantum computers with HPC, the integration of analogue quantum computers remains relatively unexplored. This paper aims to bridge this gap by contributing in the following way:

\emph{Comprehensive Survey}: We conduct a comprehensive survey of existing quantum software tools with analogue capabilities.

\emph{Readiness Assessment}: We introduce a classification and rating system to assess the readiness of these tools for HPC integration.

\emph{Gap Identification and Recommendations}: We identify critical gaps in the landscape of analogue quantum programming models and propose actionable recommendations for future research and development.

\end{abstract}

\section{Introduction}
\label{sec:intro}

Quantum computing (QC) offers a fundamentally different computational approach with the potential to revolutionise fields such as cryptography, quantum chemistry, and material science \cite{gill_quantum_2022}. Major vendors are currently focused on scaling gate-based quantum computers \cite{noauthor_ibms_nodate, noauthor_google_nodate, noauthor_quantinuum_nodate, noauthor_quantum_nodate-1}. However, fault tolerance is essential to unlocking the full potential of these machines, and while significant progress has been made, substantial challenges remain \cite{campbell_series_2024}.

Analogue quantum computing offers a potential solution to some of the challenges gate-based quantum computers face. Recent research \cite{shi_resource-efficient_2020, liang_towards_2023, smith_programming_2022, peng_simuq_2024} suggests that these machines are promising candidates for addressing a range of practical problems, including Quantum Hamiltonian Simulation (QHS) \cite{daley_practical_2022} and combinatorial optimisation \cite{headley_approximating_2022, liang_hybrid_2022}. 
However, due to their relatively lower popularity, there is a shortage of software tools which could fully leverage potential of analogue quantum computing \cite{liang_towards_2023}. 

Regardless of type, quantum devices will not replace classical computers. Instead, they are seen as hardware accelerators, similar to GPUs, capable of boosting the computational power of high-performance computing (HPC) systems \cite{humble_quantum_2021}. For instance, quantum computers could accelerate quantum chemistry problems in the strong-correlation regime while leaving the weakly correlated interactions to classical methods \cite{bauer_quantum_2020}. While the search for specific applications is ongoing, it is clear that quantum computing will require close integration with classical processors, presenting challenges at both hardware and software levels to become effective.

In this work, we focus on the software side of things. More precisely, we review existing programming models and tools that target analogue quantum computation and compile to analogue devices. Furthermore, we assess their readiness for HPC-QC integration. Our preliminary research shows that this space is underexplored compared to gate-based quantum computing. As such, there is a need to develop new programming models and frameworks to fully exploit analogue quantum devices.

The rest of the paper is organised as follows: In \cref{sec:background}, we motivate our work and give the necessary background. Next, in \cref{sec:related}, we cover related reviews on analogue quantum computing, quantum programming models and HPC-QC integration. Then, \cref{sec:methods} outlines a taxonomy and methodology for our analysis. In \cref{sec:analysis}, we collect existing quantum software frameworks supporting analogue computation. We perform a detailed analysis of their suitability for HPC environments. We close with a discussion of our results in \cref{sec:discusion} and an outlook for future work in \cref{sec:conclusion}. 
\section{Background}
\label{sec:background}
The following section outlines the background for this work. We begin by defining high-performance computing and quantum computing. Then, we introduce the categories of analogue quantum computing, highlighting their potential advantages.
Finally, we explore the concept of programming models, their importance, and the motivation for studying them in the context of quantum computing.

\subsection{High-Performance Computing}
\label{sec:hpc}
The objective of High-Performance Computing (HPC) is to solve problems that are intractable for standard desktop machines. It is generally achieved by partitioning the studied problem and performing massively parallel computation. For years, the dominant approach to realise this was connecting many commodity processors via networks to create large-scale distributed systems. To realise data transfer, the dominant communication protocol is Message-Passing Interface (MPI) \cite{mpiforum_mpi_2023}. As such, any software tool hoping for high usage in the HPC domain must be amenable to MPI to be adopted \cite{artigues_evaluation_2019}.

However, since 2004, a new paradigm has taken hold where modern HPC systems become more complex and built of specialised devices -- so-called accelerators. One such accelerator is a general-purpose graphical processing unit (GPGPU) \cite{fan_gpu_2004}. Indeed, in 2018, 40\% of systems in the Top500 list \cite{strohmaier_top500_2024} were accelerated predominantly with GPUs \cite{artigues_evaluation_2019}. As of today, most new systems at the top of the list are accelerated. We note that accelerators are not new in the HPC space -- in the early 1980s ICL Distributed Array Processors were installed in the physics department at the University of Edinburgh, and the programming model and challenges were remarkably similar to that of modern GPU architectures \cite{EPCC_history, SY86}. The total dominance of accelerated architectures however, is a recent development.

Given the variety of available hardware, a significant challenge is the portability of the software - making the same code run efficiently on different devices. Moreover, as the complexity of supercomputers grows and emerging accelerator technologies enter the market, the problem will become even more severe \cite{artigues_evaluation_2019}.

The main application of HPC is in the fields related to scientific computing, where the problems studied have computational-intensive characteristics.

In recent years, interpreted languages such as Python have become popular in scientific computing. The reason being their flexibility and increased developer productivity. However, for the most part, the dominant large-scale codes running on supercomputing systems are compiled, as native code has better performance characteristics \cite{lin_comparing_2021} and compilation offers an opportunity to detect and correct errors ahead of runtime. 
In practice, those codes should be considered the leading candidates for adding quantum routines and using quantum computers at the industrial scale. 

Summarising, the grand challenges faced by the HPC community are \emph{clarity}, \emph{productivity}, \emph{portability} and \emph{performance}. 
There is a lot of legacy baggage, and most practical quantum computing will need to be integrated with existing code rather than focusing solely on developing new applications.
Therefore, adhering to the constraints of the HPC field is crucial for emerging technologies and software tools.
\subsection{Quantum Computing}
\label{sec:qc}
The fundamental concept of quantum computing is that of a quantum bit, or \emph{qubit} for short. A qubit is a system living in two-dimensional Hilbert space, which, similar to a classical bit, has two possible states labelled $\ket{0}$ and $\ket{1}$. Contrary to the classical picture, a qubit can be in any linear combination (or superposition) of those two states. Then the state of a single qubit system is described by, $$\ket{\psi} = \alpha\ket{0} + \beta\ket{1}, \alpha, \beta \in \mathbb{C}.$$

For the system of $n$ qubits, the corresponding Hilbert space is created by taking the Kronecker tensor product of the components, producing space that has dimension $2^n$. The system's state is a unit vector embedded in this vector space. As the problem size grows exponentially, the general simulation of an $n$-qubit system is prohibitively expensive for classical computers.

\begin{figure}
    \centering
    \includegraphics[width=0.5\linewidth]{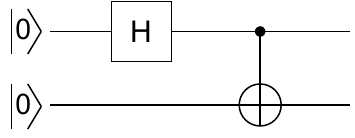}
    \caption{Quantum circuit preparing Bell state from zero state. Lines represent wires or qubits; square with H denotes Hadamard gate which puts qubit in superposition; followed by two-qubit controlled NOT gate. A quantum circuit diagram is a temporal description, unlike a classical electronic circuit diagram, which is spatial.}
    \label{fig:bell}
\end{figure}

Qubits in superposition of states can interfere, producing creative or destructive interference patterns. Upon measurement, by leveraging interference, desired states can be amplified while suppressing the others. This phenomenon is another source of the promised power coming from quantum computing.

Furthermore, qubits can be \emph{entangled}. Entanglement is a concept without classical analogue, where the qubits are correlated with each other and cannot be represented as a direct product of single-qubit states. This results in operations on one qubit affecting all others with which it is entangled.

In quantum mechanical formalism, the quantum system's state evolves in time as a unitary transformation. The unitary transformation ensures that the state is normalised. In the standard circuit-based model of quantum computation, unitary transformations are building blocks of the program and are referred to as \emph{quantum gates}. Quantum circuits have a visual representation (see \cref{fig:bell}), which simplifies reasoning about the quantum algorithms.

The quantum complexity advantage is achieved by combining the superposition, entanglement and constructive/destructive interference of qubits.
\subsection{Analogue QC}
\label{sec:analog-qc}
Fundamentally, any real-world implementation of a quantum computer is an analogue machine.
For example, in neutral-atom Rydberg-interaction machines, the qubit is implemented by a neutral atom, with the two distinct quantum states being the atom's energy levels. Then, operations applied to qubits are performed by the inherently continuous laser pulses. Continuing with this example machine, we achieve a digital or gate-based quantum computer by specifically tuning or calibrating the allowed pulses, yielding a discrete basis gate set. 

On the other hand, by allowing ourselves to control the device arbitrarily with continuously parameterised pulses, we stay at the analogue level. Operation on the analogue level provides an advantage by giving access to operations native to the device. By taking fewer steps, we can achieve the same results with higher efficiency in specific scenarios. 
Analogue QC can be roughly split into three categories: Analogue Hamiltonian Simulation (AHS), Adiabatic Quantum Computing (AQC) and Quantum Annealing (QA).


In the literature on analogue quantum computing, there is also a discussion of analogue pulses. While serving as a common interface to programming the quantum device (e.g. Bloqade \cite{quera_bloqade_2023} or Pulser\cite{silverio_pulser_2022}),
many packages in the field implement either limited Hamiltonians or are not flexible enough for arbitrary AHS (for example, Qiskit Pulse \cite{alexander_qiskit_2020} in general lacked global addressing to realise fully analogue system evolution). In cases where limited analogue pulses serve as a reinforcing technique to standard gate-based quantum computing, these are referred to as semi-analogue quantum programming \cite{smith_programming_2022}.

\subsection{Analogue Hamiltonian Simulation}
\label{sec:ahs}
Given the many-body Hamiltonian $H=\sum{H_{i,j}}$, where $H_{i,j}$ describes the local interactions between system components $i, j$, we want to simulate the system evolution for time $t$ given by $e^{-iHt}$.
The standard method to approximate this evolution in the gate-based formalism is via the Lie-Trotter product formula \cite{trotter_product_1959}, a process often referred to as \emph{Trotterization}. The error induced by the use of Trotterization is inversely proportional to the Trotter time-step $dt$.

If we want to implement this approximation using a circuit model with a realistic, fixed gate set, we face a few challenges. As summarised in \cite{clinton_hamiltonian_2021}, we can generally assume that implementing an arbitrary term in the Trotterized product requires at least one two-qubit gate. Two-qubit gates generally take longer to physically implement and this overhead places a lower bound on the achievable Trotter time-step $dt$ regarding the number of operations required and the device's capabilities to switch on and off required controls. Also, for smaller $dt$, it is hard to implement gates with high fidelity, further increasing errors in the computation.

However, rather than implementing this evolution with quantum circuits, we can directly program Hamiltonians of the analogue quantum computer (another term used in literature for the quantum device in this specific use case is \emph{analogue quantum simulator}) and synthesise desired target Hamiltonians \cite{peng_simuq_2024}. One avoids overheads associated with the gate-based implementation and implements Hamiltonian evolution with a shorter pulse duration. Moreover, expensive interactions in gate-based formalism (multi-qubit gates) can be applied globally \cite{daley_practical_2022}.

Continuing with the previous example, interaction modelling Trotterized term is implemented natively and executed for time $dt$. Therefore, if we compare circuit-based runtime to Hamiltonian-oriented analogue runtime, the simulation takes $O(t/dt)$ vs $O(t)$ respectively \cite{clinton_hamiltonian_2021}. 
Based on the previous discussion, the scheme achieves better time-to-solution. Moreover, one can expect slower error buildup from the number of operations used \cite{clinton_hamiltonian_2021}.

The natural limitation of this approach is that the underlying Hamiltonian of the studied system must map to the device Hamiltonian (i.e. Hamiltonian, which describes the possible evolution of the qubits controlled by the device) \cite{daley_practical_2022}. The problem of mapping a device Hamiltonian to an arbitrary problem Hamiltonians is studied by the optimal control community. For general Hamiltonians, it is a hard problem \cite{peng_simuq_2024}. This challenge implies that doing analogue Hamiltonian Simulation is a highly specialised problem.

One might fairly point out that the analogue computation has significant noise and error resilience difficulties. This bottleneck is true both in the classical and quantum case. Nevertheless, in the Noisy Intermediate-Scale Quantum (NISQ) era \cite{preskill_quantum_2018}, where noise and error rates are limiting parts of QC, and devices suffer from a lack of scarce resources for overheads caused by error correction, the centre of the stage is taken by the algorithms that are resilient to noise, such as variational quantum eigensolver (VQE) \cite{bharti_noisy_2022}. In those cases, we can get away with using analogue devices \cite{clinton_hamiltonian_2021}.

Indeed, a growing body of science shows the usefulness of performing pulse-level quantum computation (provided the pulses can be sufficiently well controlled) not only in Quantum Hamiltonian Simulation \cite{daley_practical_2022} already mentioned but also for Quantum Approximate Optimization Algorithm (QAOA) \cite{headley_approximating_2022}, VQE \cite{liang_napa_2024}, and machine learning (ML) \cite{kornjaca_large-scale_2024}.

From the programming perspective, Hamiltonian Simulation is of interest, as the natural formalism of building operators is high-level. One can employ Hamiltonian Modeling Languages (HML), which have a low barrier to entry.
\subsection{Adiabatic Quantum Computing}
\label{sec:aqc}


A specific form of AHS, called Adiabatic Quantum Computing (AQC) is based on the Adiabatic Theorem, which roughly states that if we start in the eigenstate (e.g. ground state) of a given time-dependent Hamiltonian $H(t)$ during the evolution under the Schrödinger equation, the system will stay in the instantaneous Hamiltonian eigenstate, provided the change in $H(t)$ is ``slow enough'' \cite{albash_adiabatic_2018}.

Formally, using the definition from Aharonov \cite{aharonov_adiabatic_2007} and the definition of $k$-local Hamiltonian, i.e. Hamiltonian, which is a sum of Hamiltonians acting non-trivially on at most $k$ particles:

Adiabatic Quantum Computation is defined by two k-local Hamiltonians, $H_0$, $H_1$, which act on $n$ particles with $m \ge 2$ states. The ground state of $H_0$ is a non-degenerate product state. The output is given by defining a schedule function $s(t): [0, t_{run}] \to [0, 1]$, and $t_{run}$ such that the final state of adiabatic evolution of the $H(s) = (1 - s)H_0 + sH_1$ is $\varepsilon$-close in Euclidean norm to the ground state of $H_1$.
The algorithm implementing AQC is known as a Quantum Adiabatic Algorithm (QAA). 

The history of adiabatic quantum computing starts with an established classical algorithm - Simulated Annealing (SA) \cite{kirkpatrick_optimization_1983}. In 1988, it was established that it is possible to use ground states of quantum Hamiltonians to encode solutions to combinatorial problems \cite{apolloni_quantum_1989}. Shortly after, Apolloni et al. \cite{apolloni_numerical_1988} proposed implementing a quantum-inspired algorithm called Quantum Annealing (QA), analogous to simulated annealing. The initial idea behind QA was that instead of thermal fluctuations (used in SA), simulated quantum fluctuations and simulated quantum tunnelling were means to solve combinatorial optimisation problems. In the optimisation space landscape, with many local minima, thanks to quantum tunnelling, QA, in principle, can penetrate hills better to find the optimal solution. In those early days, QA was also known as quantum stochastic optimisation (as introduced in the original article \cite{apolloni_quantum_1989}) whereas now, the original QA executed on a classical devices is referred to as Simulated Quantum Annealing (SQA) for clarity.

In parallel, starting from 1999, another concept was evolving. While early work on quantum computing was defined in a quantum circuit model, mainly through Deutsch and Penrose \cite{deutsch_quantum_1997}, who introduced it, Farhi et al. \cite{farhi_quantum_2000, farhi_quantum_2001} realised that by leveraging the adiabatic theorem, one could map problems to the Hamiltonians and perform computation. Soon after, it was shown that AQC is a universal model of quantum computation, equivalent to circuit model \cite{aharonov_adiabatic_2007}. Moreover, there exist routines which map between those two models, with polynomial overhead only \cite{albash_adiabatic_2018}. What followed was the re-implementation of seminal quantum algorithms to their adiabatic form. Examples include: Grover's search \cite{roland_quantum_2002}, the Deutsch-Jozsa algorithm \cite{sarandy_adiabatic_2005, wei_modified_2006} and prime factorisation \cite{jiang_quantum_2018}

While initially, QA was a classical algorithm, with the progress in quantum computing and information, the experimental story for QA is slightly different. Based on the developments in the control of quantum systems and theoretical understanding of AQC, Brooke et al. \cite{brooke_quantum_1999, brooke_tunable_2001} realised experimentally a quantum annealer, i.e. a device which implements QA algorithm on a quantum hardware which approximately performs adiabatic quantum computation. Approximately is a key word here, as most AQC theory deals with closed quantum systems - systems isolated from the environment. In practice, such systems do not exist. After all, there is always the presence of noise due to thermal and electromagnetic fluctuations \cite{yarkoni_quantum_2022}. Moreover, quantum annealers implement `stoquastic'\footnote{The term `stoquastic' comes from a combination of stochastic and quantum.} AQC \cite{albash_adiabatic_2018}. That is, the Hamiltonians that are being evolved are stoquastic, which means the off-diagonal terms in the computational basis are non-positive. Hamiltonians of this form do not suffer from a sign problem and are easier to solve\footnote{Indeed, problems encoded in stoquastic Hamiltonians can be efficiently solved by quantum Monte Carlo methods.} \cite{hormozi_nonstoquastic_2017}. As such, general AQC is different from real-world stoquastic AQC implementations modelled with transverse-fields Ising Hamiltonians \cite{albash_adiabatic_2018}. Indeed, while AQC is a universal model of computation, stoquastic AQC is not \cite{albash_adiabatic_2018}. This is the root of many controversies surrounding QA hardware and its prospective advantage over classical algorithms \cite{albash_demonstration_2018, albash_adiabatic_2018, rajak_quantum_2023, yarkoni_quantum_2022}.

Due to the distinction between AQC and stoquastic AQC and the intended use case, stoquastic AQC (StoqAQC) is also known simply as Quantum Annealing in literature \cite{albash_adiabatic_2018, yarkoni_quantum_2022}. For the rest of this review, we will use terms interchangeably, favouring the term QA when describing work done on real-world quantum annealers.

\subsubsection{Applications}
Adiabatic quantum computing, being a universal model of computation, has a similar set of applications as circuit QC. As such, standard quantum algorithms are implementable in this formalism.
From the QA point of view, the main application is that of combinatorial optimisation. Devices provided by D-Wave allow for solving Binary Quadratic Models (BQM) in the form of the Ising Model and Quadratic Unconstrained Binary Optimisation (QUBO) problems \cite{yarkoni_quantum_2022}. It is generally possible to map many vital problems to this representation. Notable examples include
Maximum Independent Set (MIS) problems \cite{yarkoni_boosting_2017} or satisfiability (SAT) problems \cite{hsu_quantum_2018}.
Still, the search for useful applications continues and remains one of the significant challenges of quantum computing.

\subsubsection{Advantages Over Circuit Model}
Given existing work on the quantum circuit model and its many advantages, a fair question to ask is: ``Why bother with AQC and QA?'' Setting aside pure scientific interest in trying things differently, there are practical reasons for considering those concepts, especially in the era of NISQ devices, where there is a strong drive to demonstrate practical quantum advantage.

The first clear advantage over gate-based systems is the scaling. To date, the pioneering QA vendor, D-Wave, offers quantum annealer chips with over 5000 qubits. While those have limited connectivity, the scale is beyond any other technology, bringing us closer to the regime where a clear advantage over classical methods can be shown. Indeed, recently, King et al. \cite{king_computational_2024} have demonstrated quantum advantage in a simulation of nonequilibrium dynamics of a magnetic spin system quenched through a quantum phase transition.

Another advantage from the perspective of this work is that the programming model for quantum annealers is at a higher level than the circuit model.
Although it might be argued that QA is programmed closer to the hardware with use of pulses and qubit topologies, in AQC and AHS, in general, one can reason about the problem at the level of Hamiltonian terms, breaking away from the high granularity of gate-based paradigm.
In fact, Ocean SDK \cite{d-wave_systems_inc_ocean_2024} offered by D-Wave for solving QUBO problems is strikingly similar to classical software used for combinatorial optimisation problems. Such an interface, in turn, makes it easier to integrate quantum annealers as accelerators within existing code bases. Lowering barriers to technology adoption increases the likelihood we can tackle practical problems earlier. Moreover, both the natural applications for QA and the programming model fit well into the hybrid HPC-QC integration model, where it is relatively easy to partition problems and offload to different devices (GPU, CPU or QPU). Indeed, recently, D-Wave introduced a hybrid solver which decomposes the encoded problem and solves it on their hybrid quantum-classical cluster \cite{d-wave_systems_inc_d-wave_2020}.

One promising avenue is parallel quantum annealing \cite{rajak_quantum_2023}. Exploratory work by Pelofske et al. \cite{pelofske_parallel_2022} has shown that it is possible to solve many independent problems using QA during a single annealing process. The main observation was that while there was a slight drop in accuracy, the speed-up was considerable exhibiting good scaling for the studied problem.
From the perspective of HPC, any parallelisation strategy on emerging technologies is of interest.

The final consideration simply is that given many different quantum hardware technologies, each with its advantages and disadvantages, and without a clear winner, it is possible that the future HPC-QC integrated systems will host different, highly specialised QPUs. In a way, this would push the existing heterogeneous computing trend to the limit. While it would create its own set of challenges for interoperability and programming, it is certainly a feasible scenario for which we need to prepare.

\subsubsection{Limitations}
From the AQC point of view, given their equivalency, the limitations are shared with the quantum circuit model and QC in general. The additional limitation is that even though equivalent, the mapping between representations creates polynomial overhead. For certain quantum algorithms, this is a bottleneck for any advantage over classical algorithms. A notable example is a Grover search, which, in its circuit form, provides a quadratic complexity advantage. So, even if the mapping is possible, it might not be practical to map circuit-based algorithms to adiabatic form \cite{albash_adiabatic_2018}.

Another issue is finding a suitable evolution time or schedule. The fundamental condition for AQC and QA is that the change in $H(t)$ is slow enough. ``Slow enough'' is bounded and depends on the assumptions about the Hamiltonian. In the worst scenario, the time evolution $t$ scales as $O(1/\Delta^3)$ -- for the most general Hamiltonian \cite{jansen_bounds_2007} and $O(log(\Delta)/\Delta^2)$ -- for bounded, infinitely differentiable Hamiltonian \cite{elgart_note_2012}, where $\Delta$ is minimal energy gap in the energy (eigenvalue) spectrum or simply spectral gap \cite{albash_adiabatic_2018}.
While this provides scaling estimations, for general Hamiltonians, estimating $\Delta$ is a hard problem. Therefore, selecting a good evolution time is a challenge and a significant limitation of the method. A possible solution is to estimate it experimentally. However, at a scale beyond classical computation, this might lead to unverifiable solutions and, in specific scenarios, suppress the quantum advantage \cite{albash_adiabatic_2018}.

Due to similar reasons corresponding to the approximation of spectral gap $\Delta$, it is hard to analyse the complexity of AQC/QA algorithms and their time-to-solution. In the quantum circuit model, the algorithm's cost is its depth or number of gates. While one could use the depth of the circuit needed to simulate an adiabatic algorithm on a gate-based device, this would assume that the circuit model is fundamental \cite{albash_adiabatic_2018}; that would also skew the results as it would include polynomial overheads due to mapping between models, which would be prohibitive for detailed analysis required in the field of HPC.
Another option is to use runtime $t_{run}$. However, it requires an appropriate energy scale of the Hamiltonian to be meaningful. Using the definition of AQC with schedule functions (outlined before), Aharanov \cite{aharonov_adiabatic_2007} introduced the concept of cost for the adiabatic algorithm as $$cost = t_{run} \max_{s} ||H(s)||,$$ where $||\circ||$ is the operator norm. The $cost$ is a dimensionless quantity, which protects from making the cost of the algorithm arbitrarily small by multiplying Hamiltonian by some size-dependent factor.
In this formalism, $t_{run}$ can be compared to an equivalent circuit-based simulation of an adiabatic algorithm and $cost$ can be analogous to circuit gate count \cite{albash_adiabatic_2018}.

As outlined by Albash and Lidar \cite{albash_adiabatic_2018}, for a limited set of algorithms where spectral gap analysis can be performed, it was shown that AQC provides an advantage over classical algorithms. Those examples mainly refer to seminal quantum algorithms.
Usually, however, the spectral gap analysis does not show any advantage, the results are inconclusive, or no information is available. Indeed, in many cases, the gap during Hamiltonian evolution becomes exponentially small with the size of the system and vanishes at the phase transition, causing $t_{run} \to \infty$.

Strikingly, there are no advantage guarantees for the original application of QA -- that of combinatorial optimisation problems \cite{albash_adiabatic_2018}. Quantum computing for combinatorial optimisation is a very active area of research due to commercial interest and availability of QA hardware. Currently, the promise is that thanks to quantum tunnelling, quantum annealers can succeed in finding minima in glassy optimisation landscapes \cite{king_computational_2024}. Moreover, it was shown only recently that quantum effects such as superposition and quantum entanglement, considered the source of quantum advantage, are present in QA devices at scale \cite{burt_d-wave_2024}.

Whereas noise mitigation and error correction are a primary focus in the field of gate-based quantum computing, they are somewhat limited in QA \cite{yarkoni_quantum_2022}. However, it is clear that the impact of the noise is severe. For example, D-Wave devices are affected significantly by the environment when the runtime is longer than a couple of microseconds \cite{rajak_quantum_2023}. It is still an open question if and how to optimally mitigate errors on analogue quantum annealers \cite{yarkoni_quantum_2022}.

While for years, the standard platform for annealers was super-conducting devices, today, trapped-ions and neutral-atoms hold much promise \cite{yarkoni_quantum_2022}. Software tools for those platforms are discussed in the following sections.

Overall, AQC and QA are fields with many promises but also many challenges. For interested readers, we refer to extensive reviews \cite{albash_adiabatic_2018, yarkoni_quantum_2022}.
\subsection{Integrating Quantum Computers with HPC}
\label{sec:hpcqc}
It is established that to leverage quantum computers fully, those machines must be tightly integrated with HPC systems. 

\begin{figure}[h!]
    \centering
    \includegraphics[width=0.9\linewidth]{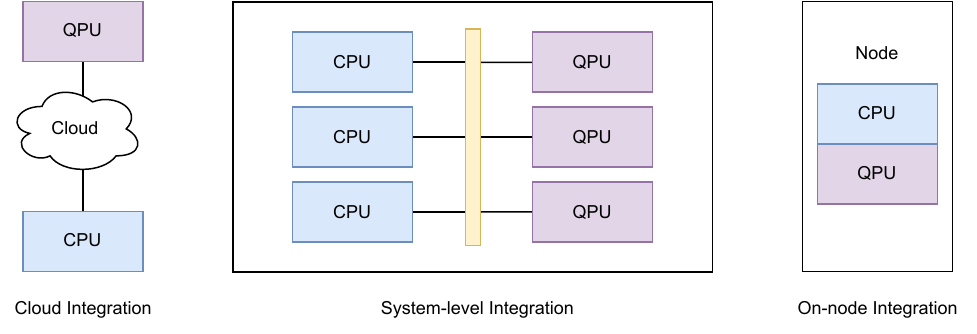}
    \caption{Proposed forms of HPC-Quantum integration. From left to right subfigures represent tighter integration with lower latency at the cost of technical challenges.}
    \label{fig:integration}
\end{figure}

How exactly such integration will look is still an open problem. Current proposed forms of integration (\cref{fig:integration}) can be roughly split into three categories.

First, there is a loose integration where logically classical and quantum systems are co-located; however, physically, access to quantum computers is given through a cloud interface. Although the easiest to realise, this setup has many bottlenecks from the HPC point of view. For many supercomputing systems, compute nodes have no access to the internet for security reasons. Moreover, offloading to the physically distant device penalises the workflow by large communication latency, which can be a bottleneck for hybrid quantum-classical algorithms.

The next step towards tighter integration is if we co-locate classical and quantum computers at the system level, within physical proximity and connect with a shared network. From the point of view of HPC, system-level integration offers dramatic growth in performance and logical coherence for quantum-accelerated scientific workflows. Nevertheless, it is significantly harder to realise it. In practice, the main challenge for such a setup would be interoperability and support for many different controls and functionalities.

Finally, we achieve the closest coupling between classical and quantum processors by performing on-node integration. It should be noted that given the sensitivity to noise of the quantum devices, on-chip integration is unrealistic. In practice, this level of integration assumes a logical hardware model where classical host shares classical information with a classical co-processor which has direct access to quantum hardware. Aside from the engineering challenges of shielding the QPU, there are additional challenges related to the synchronisation of data between host and device (common to GPU and other accelerator programming). Nevertheless, at this stage, the latency is negligible, and existing codes can be seamlessly offloaded to the quantum part, potentially resulting in significant speed-ups.

\begin{figure}[ht!]
    \centering
    \includegraphics[width=0.9\linewidth]{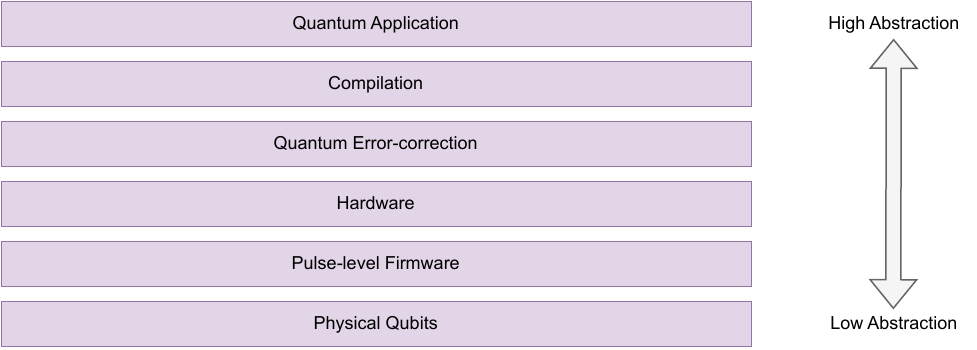}
    \caption{Layers of a quantum software stack. From top to bottom, layers proceed from the higher abstraction levels to lower, starting with quantum application layer and ending with physical qubits of the target platform.}
    \label{fig:quantum-stack}
\end{figure}

On a software level, to integrate quantum computers within the HPC ecosystem, a coordinating layer needs to be introduced which allocates and schedules computing resources for the task at hand. In general, the quantum software stack consists of multiple layers (as shown in \cref{fig:quantum-stack}): \emph{application layer, compilation layer, quantum error-correction layer, hardware layer, pulse-level firmware layer} and finally \emph{physical qubits}.

\subsubsection{Application layer}
From an HPC point of view, the role of the quantum application layer is to first implement a quantum algorithm that solves the target problem and, second, interact with established classical HPC software, for example, in a hybrid workflow. This constrains new quantum programming tools in that they need to interact with classical tools.

\subsubsection{Compilation layer}
The domain-specific application layer with high abstractions is lowered to the hardware at the compilation layer. Given the current state of things (i.e., many different platforms), compilers must simultaneously be hardware agnostic and hardware aware. Indeed, this opens room for the use of different intermediate representation (IR) dialects for different tasks \cite{mintz_qcor_2020, nguyen_quantum_2021, mccaskey_mlir_2021, guo_isq_2023, healy_design_2024}, such as general translation, general circuit optimisation, followed by hardware-specific gate optimisation \cite{schmale_backend_2022}. Elsharkawy et al. \cite{elsharkawy_integration_2023} stress the need for standardisation in this space, and indeed, there are many attempts to do so (OpenQASM 3.0 \cite{cross_openqasm_2022} and QIR \cite{geller_introducing_2020} to name two).

\subsubsection{Error-correction layer}
The next layer in the stack is responsible for the error correction. For a detailed introduction and overview of the state-of-the-art in this domain, we refer to references~\cite{gottesman_introduction_2009, roffe_quantum_2019, campbell_series_2024}. In the current age of NISQ devices, this layer is mostly absent. However, many groups work on experimental realisation of fault-tolerance and subsequent implementation of error-correcting software. From the HPC perspective, the error-correction schemes are expected to require non-negligible amounts of classical compute power \cite{kurman_benchmarking_2024, barber_real-time_2024}. This not only puts an additional constraint on the hardware integration efforts, but any error-correcting software also needs to be able to interface with classical programming models and HPC environment. From a programming model perspective the distinction is whether the user implicitly allocates and operates on logical or physical qubits. Logical qubits are assumed to be error corrected, and the programming model may elide the details of the error correction routine. On the other hand the user may choose to work with physical qubits, implementing their own error correction.

\subsubsection{Hardware layer}
At the hardware layer, the operating system environment allocates classical and quantum resources to execute the compiled code. Again, in the HPC setting, this layer has to interface with classical resource management systems (for example, SLURM \cite{yoo_slurm_2003}), as it is responsible for the proper transpilation and mapping of the code onto the device. Moreover, this mapping has to be dynamic because more than one job can be scheduled onto a given device. Moreover, during the execution, the hardware layer is responsible for acquiring and communicating the results of mid-circuit measurements, which are needed for error correction and some quantum algorithms.

\subsubsection{Pulse layer}
From the software point of view, the final pulse control layer is responsible for executing the compiled quantum program according to the constraints of the actual quantum device. Here, the device processes operations based on its Instruction Set Architecture (ISA). For the most part, this layer is specific to the backend and hidden from the standard quantum software user. However, with evidence that such fine-grained control can be beneficial in the NISQ era \cite{lamata_digital-analog_2018}, some software providers decided to introduce pulse-level control \cite{alexander_qiskit_2020, cross_openqasm_2022, efthymiou_qibolab_2024, bergholm_pennylane_2022}. In the space of analogue quantum computing, on the other hand, this is the layer with which the user has to interact, creating a barrier to entry for users without a specific platform background.

To date, most work on integrating HPC-QC systems, especially on the software level, has been focused on the quantum circuit model. Aside from the work by D-Wave on their hybrid solver, the backend details of which are private, the space of analogue quantum computing is unexplored in this regard.

\begin{figure}[h!]
    \centering
    \includegraphics[width=0.75\linewidth]{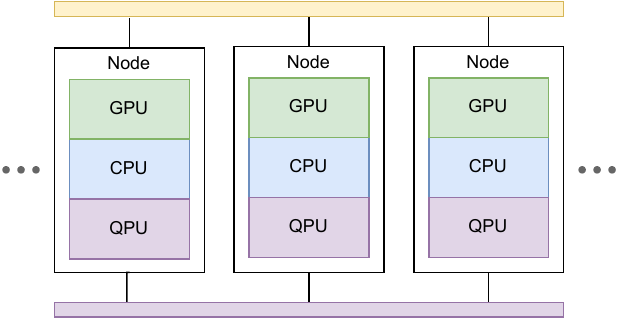}
    \caption{Example HPC architecture with integrated QPU. Stack of GPU, CPU and QPU represents a computing node; Yellow lines connecting nodes are classical interconnects; Purple lines connecting nodes are quantum interconnects used for quantum communication.}
    \label{fig:hpc-qc1}
\end{figure}

\begin{figure}[ht!]
    \centering
    \includegraphics[width=0.75\linewidth]{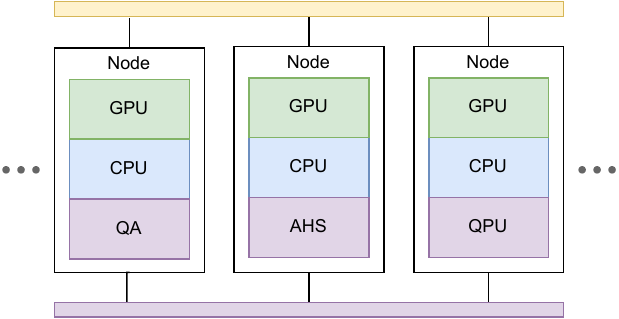}
    \caption{Example of a more specialised heterogeneous HPC architecture with different types of QPUs. All connections retain meaning from figure \ref{fig:hpc-qc1}. QA refers to Quantum Annealer; AHS to Analogue Hamiltonian Simulator; QPU to gate-based quantum machine.}
    \label{fig:hpc-qc2}
\end{figure}

Summarising, the expected architecture within the HPC system is that QPU will be an accelerator attached to a heterogeneous node (\cref{fig:hpc-qc1}).
As a matter of fact, given the specialisations and strengths of different quantum platforms, it is plausible that the actual architecture will resemble something like the one presented in \cref{fig:hpc-qc2}. Moreover, across the software stack, there is a need for tight integration between quantum device components and classical software at every stage.

To tackle real-world challenges, frameworks for programming quantum devices must support at least classical and, preferably, quantum communication. However, at the current stage, quantum communication between compute platforms is beyond our reach, and as such, we don't consider it in this work. For a review of quantum communication techniques and progress, we refer to reference~\cite{barral_review_2024}.
\subsection{Programming Models}
\label{sec:pm}

The concept of a \emph{programming models} is heavily used in the field of HPC \cite{kessler_models_2007}. There, it is usually known as \emph{parallel programming model} to hint that it refers to supercomputing parallel systems.
But what is a programming model? For the purpose of this paper, based on \cite{noauthor_introduction_nodate}, we define it as an abstraction of hardware architecture and memory. The programming model creates a framework which allows us to reason about the implementations of algorithms and their place in a program. In a way, the programming model allows for standardising how software communicates with hardware. This approach has a twofold advantage:
\begin{enumerate}
    \item Software frameworks can implement existing programming models to provide constructs for common use cases on a given hardware.
    \item Hardware producers can develop machines supporting a given programming model. This leads to better software portability.
\end{enumerate}
As such, programming models have a great impact on new hardware and software. It is crucial to have efficient, powerful, and high-level programming models for any new technology. Below, we outline motivation from the classical computing point of view and quantum computing point of view, respectively, to better illustrate the importance of developing programming models.

\subsubsection{Programming models in classical computation}
\label{sec:pmc}
As the traditional HPC requires efficient resource use, there have been many developments in the programming models space over the years, in tandem with hardware development \cite{artigues_evaluation_2019}.
The two most common parallel programming models in HPC are the \emph{Shared-memory} model and the \emph{Distributed-memory} model. In the shared-memory model, all the workers have a uniform view of memory and can read and write to it freely. In the distributed-memory model, workers have a local memory pool and need to communicate with other workers via networks. Other popular models are the \emph{Kernel-based} or \emph{Host-Device} model, for example, employed by CUDA \cite{nickolls_scalable_2008} to offload work to GPUs, and the \emph{Hybrid} model, where approaches from remaining models are intermixed for maximal performance.

Besides the hardware advances, it can be argued that the significant breakthroughs and continuous scaling can be attributed to the use and development of appropriate programming models for the task, software, and hardware at hand. Seeing the impact of those on classical computation, it is clear that similar work needs to be done for quantum computing to fully enable the technology and fulfil its promises in a practical setting.

Designing good programming models is difficult without a clear, optimal solution. Programming model creators need to balance three major factors. First, the programming model needs to be flexible enough to allow for the development of novel scientific applications. Second, it should be relatively easy so new users can quickly become productive. Finally, it should be constrained enough to penalise bad design and produce performant solutions.
\subsubsection{Programming models in quantum computation}
\label{sec:pmq}
In the current age of NISQ devices, the main quest is showing quantum advantage. Although several groups have demonstrated quantum supremacy in recent years \cite{arute_quantum_2019, king_computational_2024}, the focus has shifted to practical applications \cite{herrmann_quantum_2023}. At the research stage, the quantum programming models' requirements differ from those of their classical counterparts. As an emerging technology in the early stages the emphasis is put on low-level control so the experimenters can explore all the device's capabilities. Furthermore, performance considerations could be set aside in return for increased productivity and ease of usage for the scientists.

Nevertheless, it is desirable that when the transition towards the software engineering field happens, this transition will be as painless as possible. Therefore, it is important to consider performance, ease of use for non-domain experts, and interoperability with other classical programming models. After all, we need to consider that with high probability, quantum chemistry on a quantum computer code will run next to, or from within, Fortran code. 

Therefore, the objectives for both existing and new quantum programming models should encompass:
\begin{itemize}
    \item Support for existing hardware - so the model can be quickly adopted.
    \item Multiple levels of abstraction - so that domain experts can experiment and have high control over the device but also new users can quickly become productive
    \item Performance considerations for existing and future hardware (including classical processors), especially in spaces known to be bottlenecks. For example,
    a performant programming model could focus on limiting data movement, limiting communication overheads for future distributed systems or including support for hybrid programming models.
\end{itemize}

Given those objectives, there is motivation, both from classical and quantum points of view, to develop existing and new programming models.

\section{Related Work}
\label{sec:related}

Several existing reviews cover three subfields of quantum computing: Adiabatic Quantum Computing, HPC-QC integration, and Quantum Programming Models. 
Most of those target their domain solely, however, and to the best of the author's knowledge, there is no work analysing analogue QC programming models and existing software from the HPC-QC integration point of view.

Gill et al. \cite{gill_quantum_2022} perform a detailed comparative analysis of existing quantum software tools. By providing taxonomy and an extensive literature review, they outline existing challenges and gaps in knowledge that need to be overcome to realise quantum advantage. Additionally, they consider future directions and futuristic applications such as post-quantum cryptography.

Similarly, Serrano et al. \cite{serrano_quantum_2022} analyse existing software and hardware platform capabilities. They look at the quantum software development frameworks, compilers, simulators and error-correction tools. Resch and Karpuzcu \cite{resch_quantum_2019} provide an overview of quantum computing technology, requirements to enter the ``quantum-era'', and turning to requirements for scalability and future outlooks.

Fingerhuth et al. \cite{fingerhuth_open_2018} analysed open-source software in quantum computing, looking at the documentation, support channels, availability of tutorials, number of contributions and code readability metrics, among others.

Works by Heim et al. \cite{heim_quantum_2020} and by Garhwal et al. \cite{garhwal_quantum_2021} provide extensive review of existing programming languages targeting quantum computation, while Chong et al. \cite{chong_programming_2017} outlines requirements of a quantum compiler design and capabilities of existing compilers.

In the Adiabatic Quantum Computing and Quantum Annealing fields, Albash and Lidar \cite{albash_adiabatic_2018} provide an extensive introduction to and review of AQC. Rajak et al. \cite{rajak_quantum_2023} cover Quantum Annealing. In \cite{hauke_perspectives_2020}, authors look at the experimental realisation of QA and its real-world usage. They highlight methods for achieving large-scale QA, motivating future research.
In analogue quantum computation, Daley et al. \cite{daley_practical_2022} outlined the prospective source of quantum advantage in Quantum Hamiltonian Simulation. Clinton et al. \cite{clinton_hamiltonian_2021} cover existing and promising algorithms that could be implemented with pulses on near-term hardware. Ella et al. \cite{ella_quantum-classical_2023} cover quantum programming at pulse level, quantum-classical processing and possible benchmarks for quantum controllers.

Regarding HPC-QC integration, Humble et al. \cite{humble_quantum_2021} give perspectives on how QC-accelerated HPC systems could look. Recent work by Elsharkawy et al. \cite{elsharkawy_integration_2023} provides an extensive review of quantum frameworks, creating a blueprint for assessing their suitability for meaningful integration with classical supercomputing systems. Their work serves as a direct inspiration for this manuscript.

A review by Au-Yeung et al. \cite{au-yeung_quantum_2024} provides an excellent summary of the challenges of enhancing HPC with QC from the application point of view.
They introduce and analyse thoroughly existing quantum algorithms, discussing examples from quantum chemistry electronic structure problem and computational fluid dynamics (CFD). Importantly, the review considers all existing paradigms of quantum computing, including digital and analogue approaches. Finally, the paper identifies the challenges of interfacing quantum codes within the HPC context. They conclude that the main obstacles requiring addressing are mismatched clock speeds between the host and device, quantum data en-/de-coding, and data interconnections (with associated overheads).

A recent paper by Kaya et al. \cite{kaya_software_2024} contributed an important step towards HPC-QC integration. The group introduced the Munich Quantum Software Stack (MQSS) as a framework for integrating disaggregated quantum accelerators within the HPC system. Their work focuses on compilation, job scheduling and optimisation in an HPC-centric way. As a part of the stack, they introduce a Quantum device management interface (QDMI), which allows adding any type and number of QPUs. Additionally, Quantum Resource Manager (QRM) handles target agnostic/specific optimisation, physical fitting onto the device and scheduling. Overall, the paper fills many gaps in the field.

In a similar spirit, Beck et al. \cite{beck_integrating_2024} provide a summary of work and efforts over the past five years on the topic of HPC-QC integration. They introduce another hardware-agnostic framework for such integration, realised as a suite of software tools for resource management, offloading, and hybrid quantum-classical application deployment.

Mohseni et al. \cite{mohseni_how_2024} released an extensive review outlining challenges, outlooks and proposed solutions to build a quantum supercomputer. From the HPC integration point of view, some of the highlights include a new quantum API as an extension to HPE Cray Programming Environment (CPE), which supports languages typical to HPC (C, C++, Fortran); A plugin for Slurm that allows for quantum resource management; and finally, adaptive circuit knitting for better, distributed scaling.

Those recent, extensive reviews highlight that the field shifted to a serious effort of integrating quantum computers with HPC systems. However, neither paper looks at nor discusses in detail analogue quantum computing (QDMI, in principle, allows the addition of analogue quantum devices to the environment. However, the programming tools are predominantly gate-based).
\section{Software Rating Criteria}
\label{sec:methods}

When considering the existing software frameworks and tools, the distinctions in employed programming models are more nuanced than simple categorisation. Because of the inherently different nature of quantum computation, programming models were an essential part of theoretical and practical development from the beginning of quantum computing technology \cite{garhwal_quantum_2021}. Over the years, many programming libraries and frameworks have arisen, sometimes with great variety \cite{serrano_quantum_2022}. As such, there are many different ways to classify existing solutions. Indeed, other than focusing on the abstractions of the memory, hardware, and software executed, we can consider things like level of abstraction, employed programming paradigm and more. Below, we outline additional categories used for framework assessment in this paper, thus creating a taxonomy for the following analysis.

Similarly to Elsharkawy et al. \cite{elsharkawy_integration_2023}, we assess the viability of the quantum software from the HPC-QC integration point of view; as a result there is an overlap with the categories selected for our analysis.

\subsection{Language Characteristics}
\label{sec:paradigm}
The central factors that influence programming models and programming languages alike are their programming paradigm, syntax, and semantics.
For example, paradigm affects what a program should do or how it should be done. Likewise, the data and behaviour of the program can be organised in either object-oriented (OOP)
or functional manner. Finally, the control flow can be executed in procedural or event-driven ways, to name a few.
Decisions at this level affect the general interaction. Moreover, if the programming model is a Domain-Specific Language (DSL), the selected syntax and semantics affect the readability, conciseness, and expressiveness of the resulting model, each of which, in turn, further influences the developer's productivity.

As is usually the case, there are no right or wrong choices, only tradeoffs.
Both OOP and Functional programming have their strengths and weaknesses.
Therefore, from the standpoint of this work, we rate the software libraries based on their productivity and how well they fit within a typical HPC context.

We present the table \ref{tab:paradigm} with rating definitions for the language characteristics category.

\begin{table}[h!]
    \centering
    \begin{tabular}{|c|c|}
    \hline
        Description & Rating \\
        \hline
        Choice of programming paradigm inhibits the productivity & 1 \\
        Additionally, the syntax is inconsistent, making the programming & \\
        model hard to read and understand. & \\
        \hline
        Programming paradigm choice is challenging for the HPC domain & 2 \\
        The general design is readable but with some ambiguity. & \\
        \hline
        Choice of programming paradigm and syntax neither inhibits & 3 \\
        nor helps significantly in solving domain problems. & \\
        \hline
        Selected paradigm and syntax make the software & 4 \\
        easy to understand and provides flexibility to solve & \\
        wide range of problems. & \\
        \hline
        Exceptional choice of paradigm leads to solutions that & 5 \\
        are highly performant, extensible and easy to integrate& \\
        within the HPC ecosystem. &\\
    \hline
    \end{tabular}
    \caption{Rating scheme for software tool language characteristics.}
    \label{tab:paradigm}
\end{table}

\subsection{Software Performance}
\label{sec:performance}

Borrowing from work by Elsharkawy et al. \cite{elsharkawy_integration_2023}, we also consider the software scalability and performance as a metric. 
In a traditional HPC space, given that the current supercomputing systems are usually large-scale distributed clusters, the capability to scale with minimal overheads dictates the viability of a given software tool. 
In quantum computing, the primary consideration is given to hardware limitations and scalability regarding the number of qubits, their connectivity, and the maximum circuit depth that can be executed on the device. However, while initially not a priority, the actual software scalability becomes of growing importance. At the level of developing quantum programs, mapping arbitrary gates and circuits to corresponding instruction sets and topologies supported by devices is a challenging problem.
To achieve high software scalability
few components need to be balanced. These include:
compilation times, use of quantum resources, and communication between hardware components.

Below, in table \ref{tab:performance} we outline the rating for assessing the scalability and performance of the software.

\begin{table}[ht!]
    \centering
    \begin{tabular}{|c|c|}
    \hline
        Description & Rating \\
        \hline
        Lack of optimisations leads to inefficient, slow execution& 1\\
        and high resource consumption. & \\
        \hline
        Some consideration was given to the optimisation of & 2 \\
        quantum resources, however, there are noticeable bottlenecks. & \\
        \hline
        With some optimisation of quantum resources, the & 3 \\
        programming model meets basic requirements, but & \\
        has scalability limitations & \\
        \hline
        Programming model and compilation are performant & 4 \\
        with reasonable resource usage &  \\
        \hline
        The programming model provides exceptional performance, & 5 \\
        is highly optimised and incurs minimal latency between & \\
        classical and quantum co-processor. & \\
    \hline
    \end{tabular}
    \caption{Rating scheme for programming model scalability and performance.}
    \label{tab:performance}
\end{table}
\subsection{Tools and Ecosystem}
\label{sec:tooling}
It is often the case that the adoption of a framework by the community mimics the snowball effect. That is, the more popular the framework is, the faster the community of users grows. To start this process, the programming model must have an ecosystem of tools that make it easy to develop new code and integrate this code within existing codebases. The typical considerations include: Do development tools exist for the model? Is there a set of libraries supporting the development? Is there an active community which can help the newcomers? 

Moreover, in fields that require extensive domain knowledge, such as quantum computing, specialised tools are needed, such as those that allow the visualisation of results and inspection of the system. From the HPC point of view, specialised profilers that support analysis of parallel execution are further important. A good programming model should provide those tools or easily integrate with existing solutions.

In table \ref{tab:tooling} we outlined the rating around the tooling for a given programming model: 

\begin{table}[h!]
    \centering
    \begin{tabular}{|c|c|}
    \hline
        Description & Rating \\
        \hline
        Limited or no development tools available, small and & 1 \\
        inactive community &  \\
        \hline
        Basic development tools such as simple debugger or inspector & 2 \\
        Small but moderately active community. & \\
        \hline
        Adequate development tools for common use cases. & 3 \\
        Stable, active community. & \\
        \hline
        Strong development tools with support for debuggers & 4 \\ 
        and profilers. Additional support for basic specialised & \\
        tools (e.g. visualisation). Large and highly active community. &  \\
        \hline
        An exceptional ecosystem with a set of specialised libraries & 5 \\
        making the framework easy to adopt and use for a wide array of problems. & \\
    \hline
    \end{tabular}
    \caption{Rating scheme for programming model tooling and development ecosystem.}
    \label{tab:tooling}
\end{table}

\subsection{Applications Suitability}
\label{sec:applications}
As the main objective in the NISQ era is to show quantum utility, it is, therefore, crucial that the quantum toolchain makes it easy to implement potential candidates. If so, what is the family of applications that are easily implementable? Likewise, given the continuous search for viable applications, a good framework should be robust enough to accommodate future changes and novelty coming in the post-NISQ era.

When analysing analogue quantum computing software, we look at its target applications and capability to truly gauge its specific domain. Below, in table \ref{tab:applications}, we outline the rating system for this category.

\begin{table}[h!]
    \centering
    \begin{tabular}{|c|c|}
    \hline
        Description & Rating \\
        \hline
        Supports only generic operations at the low level & 1 \\ 
        (gates, pulses). No domain-specific structure. & \\
        \hline
        Framework has higher level constructs, making some & 2 \\ 
        domain easier to implement. & \\
        \hline
        Framework has a well-defined set of applications. & 3 \\
        \hline
        The framework implements a wide range of algorithms  & 4 \\
        targeting well defined domain (e.g. combinatorial optimisation) & \\
        \hline
        As 4 but also supports algorithms beyond projected use-cases. & 5 \\
    \hline
    \end{tabular}
    \caption{Rating scheme for programming model suitability for target application.}
    \label{tab:applications}
\end{table}

Given the natural set of applications for pulse-level analogue quantum computing, we will search for constructs that allow for easier Quantum Hamiltonian Simulation, Adiabatic evolution, and combinatorial optimisation.
\subsection{Learning Curve}
\label{sec:lcurve}
Another critical factor for the programming model's usability and adoption 
is the learning curve. Two primary considerations revolve around how easy it is to learn and be productive in a given framework and how complex the basic concepts and constructs in the model are.
It should be noted that the learning curve of the programming model is not only due to the design choices but can also be due to the underlying domain in which the given framework operates (here, quantum computing). Nevertheless, the creators of the framework should put effort into helping the user learn by documenting and providing tutorials.

Table \ref{tab:lcurve} outlines the rating categories for the programming model complexity.
\begin{table}[h!]
    \centering
    \begin{tabular}{|c|c|}
    \hline
        Description & Rating \\
        \hline
        The programming model is extremely difficult to learn & 1 \\ 
        either due to complex concepts or lack of clarity.&  \\ 
        No documentation or code examples. & \\
        \hline
        The framework is difficult to learn, requiring effort & 2 \\
        and extensive experience. Documentation and examples & \\
        are lacking. & \\ 
        \hline
        Neither easy nor hard to learn. Documentation is & 3 \\
        complete with some examples provided. & \\
        \hline
        The programming model is easy to learn with clear & 4 \\ 
        documentation and examples describing some corner cases. & \\
        \hline
        All effort has been made, so the framework is & 5\\
        easy to learn, even for beginners. Documentation, & \\
        examples and tutorials make a textbook. & \\
    \hline
    \end{tabular}
    \caption{Rating scheme for the learning curve of the programming model.}
    \label{tab:lcurve}
\end{table}

\subsection{Future prospects}
\label{sec:future}
When selecting software for a new project, an important consideration is the future prospects. Is the tool becoming more popular or obsolete? Is there active development with innovation and bug fixes implemented?
Another aspect of software tools is maturity. With maturity comes experience and safety. Mature software usually has a large user base that has tested it extensively over the years, reducing the number of bugs. Additionally, due to high usage, the changes in mature frameworks are slower, more calculated, and less drastic, preserving backward compatibility. 

Because quantum computing is a relatively new field, especially on the software side, most packages are still in active development or even at the experimental stage. 
It can be expected that many of the proposed solutions will not survive. This is a natural part of the dynamic field, such as quantum computing.
However, in recent years, major vendors moved towards standardisation of their APIs; notable examples include references \cite{aleksandrowicz_qiskit_2019}, \cite{geller_introducing_2020} and \cite{svore_q_2018}. In the quantum annealing space, D-Wave, given its long presence on the market, also has a somewhat stable offering.

The categories for the software's future prospects are presented in the table \ref{tab:future}, below.

\begin{table}[h!]
    \centering
    \begin{tabular}{|c|c|}
    \hline
        Description & Rating \\
        \hline
        Limited future prospects, project unmaintained or abandoned. & 1 \\ 
        It is in the pre-alpha stage as a prototype. & \\
        \hline
        Framework has an uncertain future. It is either incomplete, & 2 \\
        alternatively, the competition offers better solutions. &  \\
        \hline
        The software tool has stable future prospects. & 3 \\
        It is likely to remain relevant but with limited growth. & \\
        \hline
        The framework has a stable release and a promising future. & 4 \\ 
        It is possibly used by one of the large companies & \\
        resulting in growing popularity and innovation. & \\
        \hline
        Excellent future prospects. Framework is adopted & 5 \\
        by many users and is tied with many major players in a field. & \\
    \hline
    \end{tabular}
    \caption{Rating scheme for the future prospects of the programming model.}
    \label{tab:future}
\end{table}

\subsection{Parallelism}
\label{sec:parallelism}
The great success of HPC comes mainly from the use of different layers of parallelism. From the programmer's point of view, whether the parallelism is handled automatically (implicit) or requires explicit control dramatically affects the workflow and productivity. Moreover, it is essential to consider the supported granularity levels of parallelism. In the classical setting, this would refer, for example, to loop parallelism or task parallelism. In a quantum setting, one could consider shot-level or operator-level parallelism, where the expectation value calculation is distributed across many QPUs. Ideally, classical and quantum parallelism should be easily intermixed.
Finally, the important consideration of parallel computation is scalability. That is, how well does the programming model scale to a large number of processors, be they CPUs or QPUs?

Considering all that, we present the rating for the parallelism employed in a programming model below, in table \ref{tab:parallelism}.

\begin{table}[h!]
    \centering
    \begin{tabular}{|c|c|}
    \hline
        Description & Rating \\
        \hline
        There is no support for parallelism. & 1\\
        \hline
        Basic support for parallelism but either limited or inefficient. & 2 \\
        \hline
        Support for classical or quantum parallelism with limited scaling. & 3 \\
        \hline
        Good support for parallelism, including quantum parallelism. & 4 \\
        Scales well to a large number of processors. &  \\
        \hline
        Exceptional support for parallelism on different levels, & 5\\
        leveraging quantum and classical resources. HPC ready. & \\
    \hline
    \end{tabular}
    \caption{Rating scheme for the parallelism capabilities of the programming model.}
    \label{tab:parallelism}
\end{table}
\subsection{Data Management}
\label{sec:data-mng}

Another aspect affecting the programming model is the data management or simply the memory model. The ubiquitous memory models in the HPC field are the shared-memory and distributed-memory models. In quantum computing, the standard model is the shared-memory model, where the CPU and QPU communicate using shared buffers and registers. In this setup, the classical processor can pass data to the register, then the quantum device performs a quantum program, and finally, the CPU retrieves the results. While this model is straightforward, there is limited consideration about who handles synchronisation or if multiple CPUs/QPUs can access the same buffer, opening possibilities for race conditions. Moreover, due to overheads from measurements and data loading onto the QPU, communicating data from quantum devices via classical channels will generally be a more considerable bottleneck than in the current HPC world. While there is ongoing research work on distributed quantum computing (summarised well in \cite{barral_review_2024}), existing solutions are not mature enough to be directly integrated into existing quantum software frameworks \cite{elsharkawy_integration_2023}.\\

Here, in table \ref{tab:data-mng} we define rating for data management, borrowing from \cite{elsharkawy_integration_2023}.

\begin{table}[h]
    \centering
    \begin{tabular}{|c|c|}
    \hline
        Description & Rating \\
        \hline
        There are no memory model considerations. & 1\\
        \hline
        Framework supports shared memory model. & 2 \\
        \hline
        The programming model supports distributed & 3 \\
        classical memory with QPU access on-node or across nodes.& \\
        \hline
        As 3 but also includes data mapping. & 4 \\
        \hline
        The programming model supports both distributed & 5 \\
        classical and quantum memory. &  \\
    \hline
    \end{tabular}
    \caption{Rating scheme for how the programming model handles data management.}
    \label{tab:data-mng}
\end{table}
\subsection{Resource Management}
\label{sec:res-mng}

In classical HPC, resource management is prevalent in that to fully utilise the available compute resources, the resources need to be scheduled and allocated carefully. There are a few standard considerations here. For example, if a resource (e.g. GPU) can be used by two jobs simultaneously. Another consideration would be the on-node availability of the resource. Likewise, similar considerations follow since QPU can be treated as an accelerator. Elsharkawy et al. \cite{elsharkawy_integration_2023}. introduced a further concept called \emph{instruction pinning}, which considers where the instructions should be executed - on a QPU controller or the classical host. Once again, from the standpoint of the HPC user, the ideal framework would give control of the resource management on a few levels, ranging from low-level, fine-grained control to complete abstraction, with some default but unoptimised settings.

In table \ref{tab:res-mng}, similarly to data management, we work with and simplify the rating scheme from \cite{elsharkawy_integration_2023} to consider different levels of accessibility:

\begin{table}[ht]
    \centering
    \begin{tabular}{|c|c|}
    \hline
        Description & Rating \\
        \hline
        No resource management considerations. & 1\\
        \hline
        Framework supports on-node resource management. & 2 \\
        (local QPU or qubits). & \\
        \hline
        The programming model supports across-node resource & 3 \\
        management. & \\
        \hline
        The programming model additionally supports instruction pinning. & 4 \\
        \hline
        Framework offers full qubit control across the whole system. & 5 \\
    \hline
    \end{tabular}
    \caption{Rating scheme for how the programming model handles resource management.}
    \label{tab:res-mng}
\end{table}

\subsection{Hybrid Programming}
\label{sec:hybrid}
With the slowdown of Moore's Law, the HPC community trends towards heterogeneous computing. In this paradigm, compute clusters incorporate specialised hardware, different to standard CPUs. The most prevalent accelerator is GPU. It is well expected that QPU will serve as an additional accelerator in the toolkit. Therefore, any HPC-ready framework must support a hybrid or heterogeneous programming model. The framework should be aware of and operate well on the CPU-QPU interactions at the most basic level. However, it is plausible (as proposed by NVIDIA and Quantum Machines \cite{nvidia_nvidia_2024}) that other devices might be useful for quantum computation, for example, in the error-correction setting \cite{kurman_benchmarking_2024}. Finally, given the efforts in the HPC field and common practice of using different frameworks for different levels of parallelism (for example, OpenMP with MPI), one should consider if a proposed programming model is interoperable with existing hybrid parallel programming models.

In table \ref{tab:hybrid} we have outlined the rating for the hybrid programming functionality within the software programming toolkit in the table below.

\begin{table}[h!]
    \centering
    \begin{tabular}{|c|c|}
    \hline
        Description & Rating \\
        \hline
        No support for hybrid programming, limited ability & 1\\
        to leverage heterogeneous architectures & \\
        \hline
        Basic support for hybrid programming, & 2 \\
        but with limitations or inefficiencies. &  \\
        \hline
        Reasonable support for hybrid programming, & 3 \\
        but with some limitations. &  \\
        \hline
        Good support for hybrid programming, & 4 \\
        effective utilisation of heterogeneous architectures. &  \\
        \hline
        Exceptional support for hybrid programming, & 5 \\
        highly optimised for performance and scalability. & \\
    \hline
    \end{tabular}
    \caption{Rating scheme for the capability of the programming model for hybrid programming.}
    \label{tab:hybrid}
\end{table}

\subsection{Portability}
\label{sec:port}
Two common challenges in the HPC space are portability and interoperability. For the most part, current supercomputing systems have various architectures, and the hardware comes from different suppliers. While the programming model usually falls into shared-memory, distributed-memory, or GPU accelerated category, different vendors require different ways to write software targeting their hardware. This is especially evident in the accelerator space. This creates a bottleneck, as software written for one supercomputer is not necessarily performant on a different machine. Because of this, much effort has been put into developing frameworks (RAJA \cite{hornung_raja_2014}, Kokkos \cite{carter_edwards_kokkos_2014}, OpenMP \cite{openmp_architecture_review_board_openmp_2021}, OpenACC \cite{openacc-standardorg_openacc_2022}) that close this gap and support multiple platforms while abstracting hardware details from the users.

In quantum computing, it is still unclear which technology will become dominant. Each platform has its advantages and disadvantages. For superconducting qubits, this would be fast operation times and high-fidelity gates at the cost of short coherence time. Neutral atoms have excellent connectivity and long coherence times, but the application of operations is slow. Given the current state of things, it might be possible that future supercomputing systems accelerated by QPUs will use various quantum devices, each highly specialised for different needs. Therefore, it is crucial that any quantum software tools support different platforms and, from the HPC standpoint, facilitate seamless communication between them. For this paper, we have created a few categories outlined in table \ref{tab:port} below.

\begin{table}[h!]
    \centering
    \begin{tabular}{|c|c|}
    \hline
        Description & Rating \\
        \hline
        Poor portability. Difficult to port to different & 1\\
        hardware platforms or software stacks. & \\
        \hline
        Below average portability. Some limitations & 2 \\
        or challenges when porting. &  \\
        \hline
        Average portability. Can be ported to different & 3 \\
        platforms but with some effort. &  \\
        \hline
        Good portability. Can be easily ported to different & 4 \\
        hardware platforms and software stacks. &  \\
        \hline
        Exceptional portability. Highly portable and adaptable & 5 \\
        to different environments. & \\
    \hline
    \end{tabular}
    \caption{Rating scheme for programming model portability.}
    \label{tab:port}
\end{table}
\section{Analogue Programming Tools Analysis}
\label{sec:analysis}
\begin{table}
\begin{center}
\begin{tabular}{|c c c c|} 
 \hline
 
 Name & Abstraction & Host Language & Target Platform  \\ [0.5ex] 
 \hline
 Bloqade \cite{quera_bloqade_2023}      & Pulse control      & Julia/Python & NA                \\
 \hline             
 Pulser \cite{silverio_pulser_2022}         & Pulse control      & Python       & NA         \\
 \hline 
 qupulse \cite{qutech_qupulse_2024}        & Pulse control      & Python       & SC                \\
 \hline             
 Artiq \cite{bourdeauducq_artiq_2016}       & Pulse control      & Python       & TI               \\
 \hline             
 QGL \cite{bbn_qgl_2020}            & Pulse control      & Python       & SC                 \\
 \hline         
 LabOneQ \cite{zurich_laboneq_2024}        & Pulse control      & Python       & SC                 \\
 \hline
 Qua \cite{qm_qua_2024}            & Pulse control      & Python-like     & SC, NA \\
                &                       & custom language &  \\
 \hline
 OpenPulse \cite{mckay_qiskit_2018}      & Pulse control      & OpenQASM3 \cite{cross_openqasm_2022}    & SC, NA \\
 \hline
 QiskitPulse \cite{alexander_qiskit_2020}   & Pulse control      & Python       & SC \\
 \hline
 Quantify \cite{qblox_quantify_2024}       & Pulse control      & Python       & SC, NV, Spin \\
 \hline
 Q-CTRL         & Pulse control      & Python       & SC, NA, TI \\
 Boulder Opal \cite{qctrl_boulder_2024}   &                       &              & \\
 \hline
 Qibolab \cite{efthymiou_qibolab_2024}        & Pulse control      & Python/C     & Compiles to \\
                &                       &              & Qua/LabOneQ \\
 \hline
 PulseLib \cite{dalvi_graph-based_2024}      & Pulse control      & Python       & Agnostic \\
 \hline
 Amazon Braket  &  Pulse control        & Python       & QuEra NA \\
 (AHS) \cite{amazon_braket_2020}          &                       &              &           \\
 \hline
 JaqalPaw \cite{lobser_jaqalpaw_2023}       & Analogue Pulses       & Python/Jaqal & TI                 \\
                & as gates for Jaqal \cite{morrison_just_2020}    &              &                    \\  
 \hline
 Qadence \cite{seitz_qadence_2024}       & Digital-Analogue      & Python       & Compiles to Pulser \\
                & QAOA                  &              & \\
 \hline
 SimuQ \cite{peng_simuq_2024}          & HML                   & Python       & NA, DW, SC, TI    \\ 
 \hline         
 QHDOPT \cite{kushnir_qhdopt_2024}        & QUBO                  & Python       & Compiled w. SimuQ\\
 \hline
 D-Wave         & QUBO                  & Python       & DW\\
 Ocean SDK \cite{d-wave_systems_inc_ocean_2024}      & with built-in helpers   &              & \\
 \hline
 C-to-D-Wave \cite{hassan_c_2019}          & Compiles part of C    & C            & DW\\
                & to D-Wave             &              & \\
 \hline
 XACC \cite{mccaskey_xacc_2019}          & Middleware framework  & C++/Python   & SC, TI, DW \\
                & with Analogue Pulses    &              &            \\ [1ex]
 \hline
 Pennylane \cite{bergholm_pennylane_2022}      & General QC framework & Python & Agnostic \\ 
                & with Analogue Pulses& & \\[1ex]
 \hline
\end{tabular}
\end{center} 

\caption{Quantum Software with support for some form of analogue quantum computation. HML refers to Hamiltonian Modeling Language.
Platform abbreviations: NA - Neutral Atoms, SC - Superconducting, TI - Trapped Ions, DW - D-Wave Quantum Annealer, NV - Nitrogen-Vacancy Center.}
\label{tab:analog-libs}
\end{table}

Having defined the categories and metrics for the qualitative analysis, this section presents and discusses programming tools that support analogue quantum computation in more detail.
While selecting the available software tools, we faced a dilemma: if we focus strictly on the packages that allow for adiabatic quantum computing (or annealing), the list of tools is severely limited. If we relax our conditions, on the other hand, to software that enables programming at the analogue pulse level and allows for the digitalisation of adiabatic evolution, then the set of available packages is extensive and includes well-known SDKs from major vendors. After the initial review, it was clear that many software libraries provided an interface for adiabatic evolution or the quantum adiabatic algorithm (QAA).
However, most software that offers an interface for QAA is, in fact, digitising the algorithm during transpilation/compilation. This is reasonable as most existing platforms do not offer capabilities to perform true analogue adiabatic evolution. However, it creates a gap in software tooling, where the platforms with analogue capabilities are not fully utilised.

Therefore, first in \cref{tab:analog-libs}, we have collected all programming models with analogue quantum computing interfaces, be it AHS, QA or pulse level control.
We observed that, generally, there are a large number of quantum control software packages. 
While several of them offer programming at the pulse level, in general, many don't support global interactions required for the analogue Hamiltonian quantum simulation or analogue quantum adiabatic algorithm. The most common use of those is for device calibration experiments and direct system control. To date, there has been limited work done using those packages for QAA.

It is true that using a hardware simulator (e.g. Qiskit Pulse simulator), it is possible to create a setup which, in principle, can model a wide range of Hamiltonians (as shown, for example, by Greenaway et al. \cite{greenaway_analogue_2024}). However, for some platforms (e.g. transmon superconducting qubits), such a setup is currently impossible to realise in a physical setting \footnote{While preparing this manuscript, IBM introduced that Qiskit Pulse will be deprecated and Qiskit 2.0 SDK will offer fractional gates instead.}. 

As such, by analogue pulse, we understand a fully analogue pulse, with the possibility of global beams and arbitrary continuous waveforms and functions for amplitude, phase, and frequency. That is, a fully analogue pulse is capable of QAA or Hamiltonian simulation with no (or limited) Trotterization.

\begin{figure}[h!]
    \centering
    \includegraphics[width=0.95\linewidth]{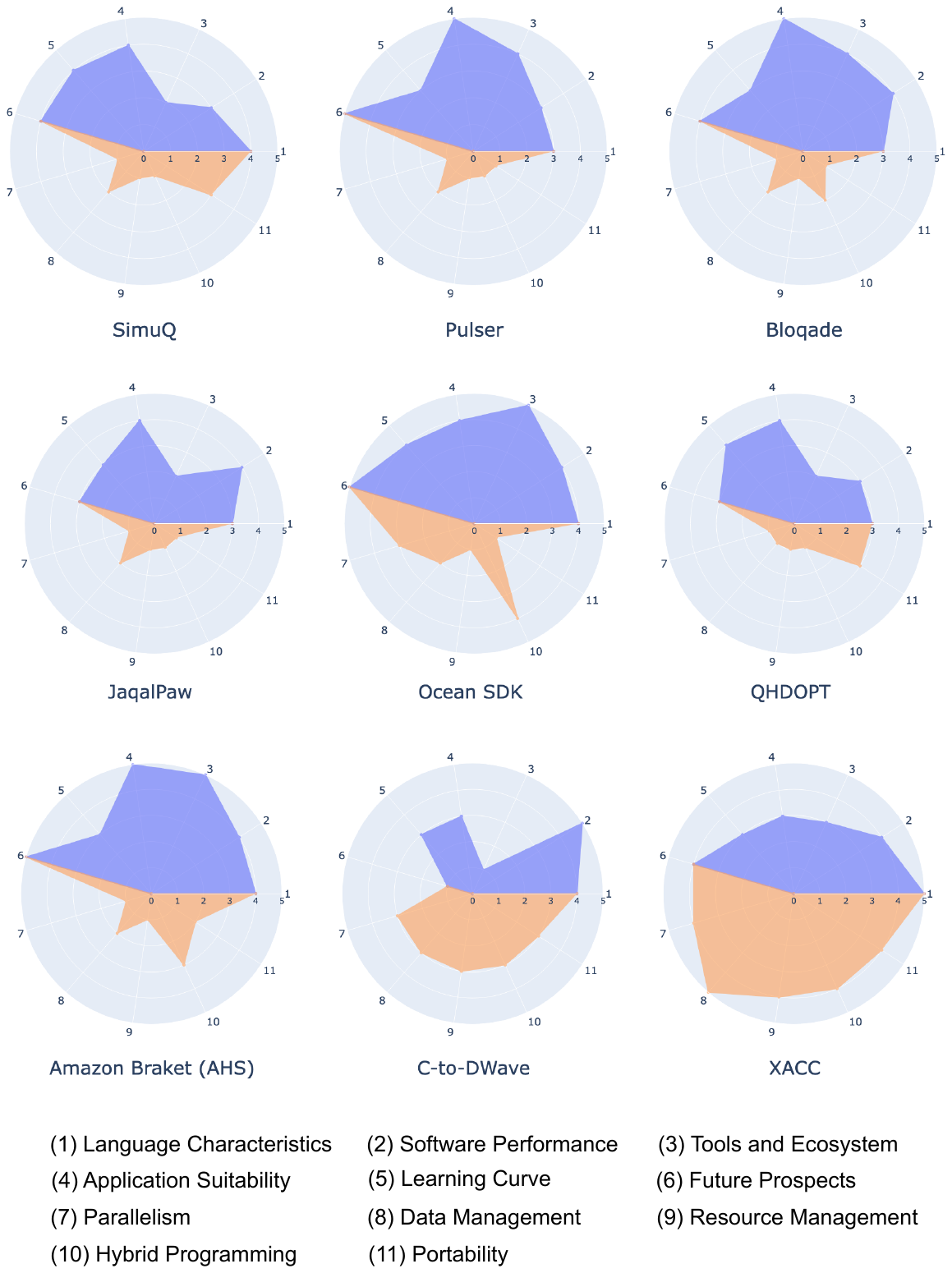}
    \caption{Radar plots representing the scores achieved by each software tool according to our rating scheme. Each number on the angular axis corresponds to categories outlined in section \ref{sec:methods}. Blue colouring covers general programming tool characteristics, whereas orange colour represents HPC-specific categories.}
    \label{fig:radars}
\end{figure}

With that in mind, for a detailed analysis, we have selected tools that either allow for QA or are capable of full AHS on existing devices. These include SimuQ \cite{peng_simuq_2024}, Pulser \cite{silverio_pulser_2022}, Bloqade \cite{quera_bloqade_2023}, JaqalPaw \cite{lobser_jaqalpaw_2023}, D-Wave Ocean SDK \cite{d-wave_systems_inc_ocean_2024}, QHDOPT \cite{kushnir_qhdopt_2024}, Amazon Braket \cite{amazon_braket_2020}, C-to-Dwave from Los Alamos National Laboratory (LANL) \cite{hassan_c_2019} and XACC \cite{mccaskey_xacc_2019}. The radar charts are presented in figure \ref{fig:radars}. Then, an exhaustive description of each model follows.

\subsection{SimuQ}
\begin{table}[h!]
    \centering
    \begin{tabular}{|c|c|}
    \hline
        Criterion & Rating \\
        \hline
        Language Characteristics    & 4 \\
        Performance                 & 3 \\
        Tooling                     & 2 \\
        Applications                & 4 \\
        Learning Curve              & 4 \\
        Future Prospects            & 4 \\
        Parallelism                 & 1 \\
        Data Management             & 2 \\
        Resource Management         & 1 \\
        Hybrid Programming          & 1 \\
        Portability                 & 3 \\
    \hline
    \end{tabular}
    \caption{Rating for the SimuQ.}
    \label{tab:simuq}
\end{table}

\label{sec:simuq}
SimuQ \cite{peng_simuq_2024} is an open-source, Python-based, domain-specific language designed for Quantum Hamiltonian Simulation. The package operates explicitly on a higher abstraction level to reduce the barrier to entry for non-experts, requiring minimal physics domain expertise. With a custom compilation framework, SimuQ targets NISQ devices from different vendors, including IBM’s Qiskit Pulse \cite{alexander_qiskit_2020} for superconducting transmon qubit devices, QuEra’s Bloqade \cite{quera_bloqade_2023} for neutral atom arrays, and quantum circuits for other general machines.

SimuQ, aside from OOP and procedural paradigm, offers a unique Hamiltonian Modelling Language (HML), which allows for accessible operator building. Those features make SimuQ readable, concise and easy to learn. Performance-wise, while their solver-based compiler includes some optimisations from the quantum resources standpoint, as authors note, there is a limited consideration given to the classical cost of compilation (although it was still more performant than Qiskit compiler at the time SimuQ was published \cite{peng_simuq_2024}). From the development perspective, there is limited tooling available. Moreover, we have found that SimuQ fails to find a solution for some Ising Hamiltonians on the QuEra platform. The application domain is well-defined, with many constructs for achieving a wide range of goals. Finally, given that the QHDOPT \cite{kushnir_qhdopt_2024} project adopted SimuQ, we deem it to have a stable future.

From an HPC perspective, SimuQ assumes a restricted Heterogeneous Quantum-Classical computation model in that the classical machine defines the quantum program and performs compilation, which takes the form of a pulse schedule that is offloaded to the quantum device. There are no explicit assumptions when it comes to data management. The framework employs an implicit shared-memory model between classical and quantum devices. SimuQ assumes local qubits (sites) allocation, and in that manner, little consideration is given to resource management. It also does not support classical communication and hybrid programming. Finally, given its compilation infrastructure, it is portable, targetting D-Wave, IBM, IonQ and QuEra and, being Python-based, it can be easily installed on most systems.
\subsection{Pulser}

\begin{table}[h!]
    \centering
    \begin{tabular}{|c|c|}
    \hline
        Criterion & Rating \\
        \hline
        Language Characteristics    & 3 \\
        Performance                 & 3 \\
        Tooling                     & 4 \\
        Applications                & 5 \\
        Learning Curve              & 3 \\
        Future Prospects            & 5 \\
        Parallelism                 & 1 \\
        Data Management             & 2 \\
        Resource Management         & 1 \\
        Hybrid Programming          & 1 \\
        Portability                 & 1 \\
    \hline
    \end{tabular}
    \caption{Rating for the Pulser.}
    \label{tab:pulser}
\end{table}

\label{sec:pulser}
Pulser \cite{silverio_pulser_2022} is a package aiming for the accessible design and execution of analogue pulse sequences, targeting neutral atoms. It is flexible enough that it should be executable on any such platform. It is embedded in Python, and a QuTip \cite{johansson_qutip_2013} simulator is included. The data structures are organised in an object-oriented manner with a sequence builder at the centre. The syntax is generally clear and readable; however, as a quantum hardware control software, it is low-level.
The execution model lets the user offload the pulse sequence onto the backend. However, there is no classical control during the execution. The user can generally select a local emulator or a remote Pasqal system in the cloud. Alternatively, one can also perform tensor network simulation remotely. From the performance point of view, certain routines (such as qubit placement) implemented in Python can be less performant than those written in a compiled language.
While some consideration is given to the optimisation of quantum resources, much is left to the backend or the user. The framework has extensive tooling, which includes a built-in emulator for verification and visualisation tools. Additionally, it has a growing community with a recent, stable release (1.0.0). While Pulser has a certain amount of higher-level constructs for more straightforward pulse definition, it does not target specific applications, e.g., a quantum adiabatic algorithm needs to be implemented from the bottom up.
One interesting feature of Pulser is that, besides hosting a library of device specifications, it allows users to define their own analogue devices and experiment freely. 
Pulser has a bright future because it is tied to Pasqal hardware, is actively developed, and serves as a base for a new digital-analogue framework, Qadence \cite{seitz_qadence_2024}.

From the HPC point of view, Pulser does not have any direct consideration for classical or quantum parallelism. The employed memory model is a shared-memory model with local resource control and no communication. There is no support for hybrid programming either. Finally, from the portability point of view, Pulser supports only the Pasqal devices through the cloud (or Microsoft Azure). Up until recently, the main intermediate representation for the pulse sequences was in a JSON format, which, while not ideal, has a potential for future portability, assuming different platforms can compile from it. On the other hand, from the classical perspective, being a Python package distributed by pip, it is portable and can be installed on most HPC systems.
\subsection{Bloqade}

\begin{table}[h!]
    \centering
    \begin{tabular}{|c|c|}
    \hline
        Criterion & Rating \\
        \hline
        Language Characteristics    & 3 \\
        Performance                 & 4 \\
        Tooling                     & 4 \\
        Applications                & 5 \\
        Learning Curve              & 3 \\
        Future Prospects            & 4 \\
        Parallelism                 & 1 \\
        Data Management             & 2 \\
        Resource Management         & 1 \\
        Hybrid Programming          & 2 \\
        Portability                 & 1 \\
    \hline
    \end{tabular}
    \caption{Rating for the Bloqade.}
    \label{tab:bloqade}
\end{table}

\label{sec:bloqade}
Bloqade \cite{quera_bloqade_2023} is a software library from QuEra targeting the design and execution of analogue quantum Hamiltonian dynamics. It targets solely neutral atom architecture. It allows for easy generation of pulses and waveforms for quantum registers with arbitrary topologies. Furthermore, the package features a fast simulation of the entire Hilbert space and its subspaces subject to the Rydberg blockade. The sequences of pulses can be further compiled and executed on QuEra quantum devices using the Amazon Braket \cite{amazon_braket_2020} service. The employed programming paradigm mixes functional capabilities with OOP constructs. In general, the API is low-level. However, it also provides higher-level features, such as creating general observables using operator expressions. This results in a clear and expressible library suitable for most neutral atom experiments. The framework is written in Julia with an additional interface in Python, making it performant and classically portable.
Additionally, simulations of the quantum systems can be offloaded to GPUs using CUDA. The overall tooling ecosystem is extensive, with built-in emulator and visualisation packages. With complete documentation and a set of examples, Bloqade is relatively easy to learn, with the main obstacle being the domain knowledge. Moreover, it has an active community, and with QuEra devices being offered by Amazon Braket service, it has a promising future. 

Looking at Bloqade from the HPC perspective, there is a form of quantum parallelism on QuEra devices, which allows task parallelism on a single device.
The assumed execution model does not allow classical control during the QPU execution. There is limited consideration of memory model and resource accessibility, with mostly shared-memory, site (qubit) access provided by the system. There is no communication consideration, be it classical or quantum. While there is no direct functionality for partitioning problems for hybrid accelerators (e.g. GPUs), the fact that the emulation part already uses CUDA means there is future potential in this space. Finally, from the quantum portability perspective, the only supported platform is a neutral atom device from QuEra. On a Braket system, the intermediate representation has the form of JSON, which might offer some potential in the future.

\subsection{JaqalPaw}

\begin{table}[h!]
    \centering
    \begin{tabular}{|c|c|}
    \hline
        Criterion & Rating \\
        \hline
        Language Characteristics    & 3 \\
        Performance                 & 4 \\
        Tooling                     & 2 \\
        Applications                & 4 \\
        Learning Curve              & 3 \\
        Future Prospects            & 3 \\
        Parallelism                 & 1 \\
        Data Management             & 2 \\
        Resource Management         & 1 \\
        Hybrid Programming          & 1 \\
        Portability                 & 1 \\
    \hline
    \end{tabular}
    \caption{Rating for the JaqalPaw.}
    \label{tab:jaqalpaw}
\end{table}

\label{sec:jaqalpaw}
JaqalPaw \cite{lobser_jaqalpaw_2023} is an offering from the Sandia National Laboratory for QSCOUT Trapped Ion devices. JaqalPaw is written as a framework in Python where the user defines custom gates in terms of pulses and waveforms. These are later imported to Jaqal assembly language format \cite{morrison_just_2020} files and freely used for further quantum processing. The gates are defined in OOP form, where a set of gates is defined as a class, and each gate definition is a method of said class. The built-in pulse definitions provide a high level of device control, which can correlate with the device's calibration data. From the QAA or Hamiltonian Quantum Simulation point of view, qubits can be manipulated via global beams, opening the possibility for truly analogue execution. The resulting programming model is straightforward, flexible and expressive but low-level, requiring domain knowledge from the user. Performance-wise, there are no direct optimisation routines. However, the resulting constructs are at the (quantum) assembly level, leaving much space for the user to implement their own optimisations.
Regarding the tooling ecosystem, a JaqalPaw emulator is shipped with the library. However, it has limited support for Jaqal's built-in multi-qubit gates (such as the Mølmer–Sørensen gate), making the development and verification of custom pulses rather challenging. From our experience, it creates a barrier to entry for users new to the field of trapped ions control. With JaqalPaw, a basic white paper is attached, explaining the structure and workflow. There are also some examples in the project repository. Overall, JaqalPaw is relatively easy for experts in the field to learn, although additional tutorials would make it easier for newcomers. The user community is relatively small; however, the project is actively updated.
Application-wise, the objective of JaqalPaw is low-level control and flexibility. It does not have a domain-specific structure (or algorithms) implemented. Overall, the JaqalPaw provides a unique solution, giving it a stable future, but given its current low popularity, it is unclear how it will grow. From the HPC point of view, there are no considerations for parallelism or communication and data management. In practice, JaqalPaw should be considered as an extension to Jaqal assembly language. Indeed, the resources are allocated via Jaqal in an on-node, shared-memory model. Additionally, neither JaqalPaw (too low level) nor Jaqal directly supports hybrid programming. Finally, portability is limited as the JaqalPaw is tied to Jaqal and QSCOUT control systems.

\subsection{D-Wave Ocean SDK}


\begin{table}[h!]
    \centering
    \begin{tabular}{|c|c|}
    \hline
        Criterion & Rating \\
        \hline
        Language Characteristics    & 4 \\
        Performance                 & 4 \\
        Tooling                     & 5 \\
        Applications                & 4 \\
        Learning Curve              & 4 \\
        Future Prospects            & 5 \\
        Parallelism                 & 3 \\
        Data Management             & 2 \\
        Resource Management         & 1 \\
        Hybrid Programming          & 4 \\
        Portability                 & 1 \\
    \hline
    \end{tabular}
    \caption{Rating for the D-Wave Ocean SDK.}
    \label{tab:dwave}
\end{table}

\label{sec:dwave}
When it comes to quantum annealing frameworks, the D-Wave offering remains the primary option. 
D-Wave Ocean software development kit (SDK) \cite{d-wave_systems_inc_ocean_2024} is a collection of open-source libraries for solving combinatorial optimisation problems. Among others, they offer tools for graph mapping, constraint compilation, and encoding of problems that fit QUBO formalism. Then, the stack offloads the selected program to one of a few backends. Those include classical simulated annealing backend, D-Wave Advantage machines via their API and a new offering - hybrid solver, which decomposes the problems at hand among available devices, including QPUs and GPUs. From the programming point of view, Ocean SDK offers a mix of programming paradigms across different levels of abstraction, making the resulting code readable and easy for developers to use. Performance-wise, the D-Wave offering is efficient, providing good mapping onto the device with reasonable resource usage. The wide array of tools and libraries provides excellent development experience and allows for detailed experimentation. Extensive documentation and a set of examples ensure that the learning curve of this programming model is relatively flat. Overall, D-Wave is a leader in the QA field, and this framework's future prospects are bright.

In terms of HPC metrics, the employed execution model only allows the problem to be offloaded to the backend. There is no classical control during the execution (or it is hidden from the user). Parallelism has limited low-level control, with the main parallel construct being \emph{TilingComposite} that tiles small problems multiple times over the structured Chimera or Pegasus samplers. However, with a hybrid solver, it is easy to partition the problem and run large tasks. It is hard to assess the memory model and resource management precisely as the backend is proprietary. However, the user has a certain degree of control via Sampler API. Based on their offering, we assume the memory model is distributed classically (due to how the hybrid solver is organised). The accessibility of resources on the quantum annealer is at the node level. The hybrid solver offers good hybrid programming that is out of the box. It also allows for defining new components, justifying the high score achieved.
Unfortunately, the code produced by the framework is not portable, as the only target is D-Wave quantum annealers.

\subsection{QHDOPT}

\begin{table}[h!]
    \centering
    \begin{tabular}{|c|c|}
    \hline
        Criterion & Rating  \\
        \hline
        Language Characteristics    & 3 \\
        Performance                 & 3 \\
        Tooling                     & 2 \\
        Applications                & 4 \\
        Learning Curve              & 4 \\
        Future Prospects            & 3 \\
        Parallelism                 & 1 \\
        Data Management             & 1 \\
        Resource Management         & 1 \\
        Hybrid Programming          & 1 \\
        Portability                 & 3 \\
    \hline
    \end{tabular}
    \caption{Rating for the QHDOPT.}
    \label{tab:qhdopt}
\end{table}

\label{sec:qhdopt}
QHDOPT \cite{kushnir_qhdopt_2024} is a new, Python-based library implementing Quantum Hamiltonian Descent \cite{leng_quantum_2023} on a set of existing quantum computers. QHDOPT's objective is to lower the barriers of entry for users from different science domains who lack expertise in quantum computing. It is built on top of SimuQ, which compiles to target systems. The general design is OOP, with clear, simple and concise constructs. The user solves the problem first by defining the Q matrix and constraints for the problem. Then, a model is created, which is then compiled and offloaded to the target backend. Given the project's main aim (being easy to use), the user's control over the computation is quite limited.
From the development perspective, there are limited available tools, and some features are incomplete. Similarly, the user community is still small. Even though the documentation is still being created, it is relatively easy to learn in line with the authors' objective. At the current stage, given the project's freshness, we assume that its future is promising. It fills a niche in a field, but it is unclear how well it is going to be adopted by the wider community.

Looking at the project from the HPC point of view, there are no direct constructs for parallelism, data, and resource management. Similarly, there is no support for communication or hybrid programming. However, the package is portable both classically and quantum-wise. It can offload to D-Wave quantum annealers and IonQ devices, and being written in Python, it can be easily installed on most supercomputing systems.
\subsection{Amazon Braket - Analog Hamiltonian Simulation}

\begin{table}[h!]
    \centering
    \begin{tabular}{|c|c|}
    \hline
        Criterion & Rating \\
        \hline
        Language Characteristics    & 4 \\
        Performance                 & 4 \\
        Tooling                     & 5 \\
        Applications                & 5 \\
        Learning Curve              & 3 \\
        Future Prospects            & 5 \\
        Parallelism                 & 1 \\
        Data Management             & 2 \\
        Resource Management         & 1 \\
        Hybrid Programming          & 3 \\
        Portability                 & 2 \\
    \hline
    \end{tabular}
    \caption{Rating for the Amazon Braket.}
    \label{tab:braket}
\end{table}

\label{sec:braket}
Amazon Braket \cite{amazon_braket_2020} is a cloud quantum computing service that offers a range of quantum machines, including Gate-based superconducting processors from Rigetti and IQM, gate-based ion-trap processors from IonQ, and Neutral atom-based quantum processors from QuEra. For the purpose of this work, we only consider the Neutral atom platform, which is capable of Analog Hamiltonian Simulation (AHS). 
Braket offers a module with the same name (AHS), which allows for defining pulses and allocating qubit registers that are later mapped to the device. The programming model is, to a large extent, similar to Bloqade. From the performance perspective, although the library is Python-based, it also offers C++ API, enabling integration within HPC workflows (although quantum resources are still accessed remotely). AHS provides excellent tooling, as found in Bloqade and more, coming directly from Braket features. The model is relatively easy to learn; however, like Bloqade, the user requires domain knowledge about the neutral atom platform. We believe that the AHS, being tied to the Amazon cloud service, has a promising future, especially given the range of devices offered.
From the HPC point of view, parallelism and resource management are handled similarly to Bloqade's. That is, the user can build an atom register (the system assumes that atoms are available), and, depending on the problem size, execute independent tasks in parallel. There is no explicit way to specify communication and manage data on the device or in the broader cluster. However, Amazon Braket offers hybrid jobs, allowing easy coupling between classical and quantum processors. Finally, the programming model has limited portability, for now, targeting only QuEra devices and Amazon cloud systems.
\subsection{C-to-D-Wave}

\begin{table}[h!]
    \centering
    \begin{tabular}{|c|c|}
    \hline
        Criterion & Rating \\
        \hline
        Language Characteristics    & 4 \\
        Performance                 & 5 \\
        Tooling                     & 1 \\
        Applications                & 3 \\
        Learning Curve              & 3 \\
        Future Prospects            & 1 \\
        Parallelism                 & 3 \\
        Data Management             & 3 \\
        Resource Management         & 3 \\
        Hybrid Programming          & 3 \\
        Portability                 & 3 \\
    \hline
    \end{tabular}
    \caption{Rating for the C-to-DWave.}
    \label{tab:lanl}
\end{table}

\label{sec:lanl}
Another, albeit experimental, set of projects targeting quantum annealing is software developed by Los Alamos National Laboratory (LANL). Starting from the bottom-up, the QMASM \cite{pakin_quantum_2016} is a macro-based assembly language for D-Wave systems. Its main purpose is to create QUBO or Ising problem formulations at a low level. Higher-level operations (e.g. gates) can be created using the \lstinline$!begin-macro$ directive. Additionally, it offers a \lstinline$!bqm_type$ directive, which can specify how to interpret weights and the strength of the interaction within the formulation. Available options are ``qubo'' and ``ising''. This allows users to express parts of the program as QUBO and parts of it as Ising-model Hamiltonians.

Next, the edif2qmasm \cite{pakin_targeting_2019} project translates hardware-description language programs - Verilog/VHD - to QMASM, allowing offloading to D-Wave annealers. The main advantages of this approach are that one can intermix classical, standard language constructs with specifications for quantum operations, and further, one has control over bit/qubit widths, fully controlling the available resources (something seen in more recent frameworks, such as OpenQASM3).  

Finally, two projects are building upon the previous two. C-to-D-Wave \cite{hassan_c_2019} compiles a small subset of C language to the form that can, in principle, run on a D-Wave annealer. In that way, one can, for example, define a C function that defines the MAXCUT problem, which is then compiled to Verilog and down to QMASM for D-Wave API consumption. The C-to-Verilog compilation uses Clang tools and hence achieves good scalability; however, no further optimisations are considered on the lower levels.
Another project is QA Prolog \cite{pakin_performing_2018}, which, again, has extended functionality to define Ising-model Hamiltonian functions. This fits well into the language combined with built-in constraint-logic programming. Because it produces classical Hamiltonians, it can, in principle, offload work to classical annealers as well. Although it has definitely interesting use cases and would be of much interest to the computer science and physics communities, from the HPC point of view, Prolog as a language has limited support and is rather niche. 

While those projects are primarily proofs-of-concept, they are interesting because the stack attempts to integrate QA offloading from higher-level classical languages (C, Prolog). Even though the abstraction level could be considered high, as the user just specifies their problem, which is then compiled and executed on the D-Wave system, overall, there is limited potential for further HPC integration. There is little to no consideration for optimisation, data and resource management. Moreover, the project seems to be discontinued, with the last commits dating back to 5 years (2019). As such, although attractive, we deem those repositories unhelpful for future HPC-QC integration.
\subsection{XACC}

\begin{table}[h!]
    \centering
    \begin{tabular}{|c|c|}
    \hline
        Criterion & Rating \\
        \hline
        Language Characteristics    & 5 \\
        Performance                 & 4 \\
        Tooling                     & 3 \\
        Applications                & 3 \\
        Learning Curve              & 3 \\
        Future Prospects            & 4 \\
        Parallelism                 & 4 \\
        Data Management             & 5 \\
        Resource Management         & 4 \\
        Hybrid Programming          & 4 \\
        Portability                 & 4 \\
    \hline
    \end{tabular}
    \caption{Rating for the XACC.}
    \label{tab:xacc}
\end{table}

\label{sec:xacc}
XACC \cite{mccaskey_xacc_2019} is a framework targeting general QC technology with an extensive feature list for HPC integration. It is built from three primary layers: front-end, middle-end and backend, each providing different functionality. Thanks to this modular, compiler-like architecture, the set of supported front-end libraries and backends is immense. The programming model employed in XACC is a host-device, or kernel-based, popular in hardware accelerator space. Here, similarly, a QPU is treated as just another accelerator. The code representation is transformed to an intermediate representation (IR) residing at the middle layer, with numerous optimisations implemented for further compilation. Finally, IR is lowered to an appropriate quantum circuit format or pulse schedule for a target backend. XACC offers many levels of abstraction, supporting standard gate-based circuit building and a number of pre-implemented quantum algorithms and circuit generators. It also supports pulse-level programming. However, it is somewhat limited in this regard, bound to the QuaC \cite{otten_quac_2016} simulator backend. Additionally, it supports offloading to the D-Wave devices, making it an extremely versatile framework for different modes of quantum computing.
From the memory model and resource management point of view, XACC supports the Message-Passing Interface (MPI), which is ubiquitous in HPC and used for large-scale distributed computing. Due to the employment of MPI and an efficient compilation stack, we also rate XACC high in the performance category. While the framework seems stable enough, we found it somewhat hard to get working on different supercomputing systems. 
Additionally, it seems the project progress stalled, with much momentum being moved to CUDA-Q \cite{kim_cuda_2023}. 

While XACC is the main contender for tight HPC-QC integration, it lacks enough capability for robust analogue quantum computation, mainly in the space of AHS. One advantage of XACC is that its architecture is extensible by modules, making it relatively future-proof. In that way, some of its shortcomings could be addressed by providing an analogue-based extension package.

\section{Discussion}
\label{sec:discusion}
Having analysed the existing programming models for analogue quantum computation, we follow with a discussion, including gathered insights and limitations of this study.

\subsection{Insights}
The emerging insight from our analysis is that by and large, the existing frameworks designed explicitly for analogue quantum computation are not ready for integration with HPC systems. This is unsurprising, as they are mainly at the experimental and development levels. Nevertheless, as there is growing evidence of the utility of quantum processing at the analogue level, it is vital to either redesign existing or introduce new software tools that make HPC integration easier. 

The main limitation of existing libraries and DSLs can be traced down to the implicit execution model. In principle, it is hard to think about classical operations on a device for AQC (aside from error correction). As such, the software usually does not consider the execution model extensively, instead using a restricted model with remote submission to a QPU.
Additional limitations include little consideration of data and resource management and scalability in terms of software. 
Furthermore, for the most part, any consideration of interaction with classical systems is limited.

XACC framework is a definite outlier here, providing varied offerings and supporting offloading to D-Wave systems and programming at the pulse level. Given its modular architecture, host language being C++ and support for MPI, it is the main contender for HPC-QC integration. Indeed, our results here are confirmed by Elsharkawy et al. \cite{elsharkawy_integration_2023}, who also analysed XACC as a part of their study. 

Another limitation of existing tools, from an HPC point of view, is the extensive use of Python as a host language. While Python provides much flexibility and facilitates productivity, it is not ideal for the constraints of high-performance systems. This problem is remedied to some extent by implementing time-critical portions of code in compiled languages. An additional benefit is that Python has a low barrier to entry and is suitable for writing new scientific code. However, the actual challenge in the HPC community is adopting and extending existing code bases, which are primarily written in C and Fortran. Therefore, any HPC-worthy software toolkit has to have (at least) a C API that can be easily integrated with existing codebases.

In the QA space, D-Wave's offering reigns supreme. This can mainly be attributed to the fact that D-Wave was the first and leading vendor of quantum annealers, while other major players focused on developing digital quantum computers. Nevertheless, with the growing popularity and realisation of neutral atom and trapped ion systems, for which it is relatively easy to switch modes of operation (between digital and analogue), there is a need for open-source tools that will make it easy to program quantum annealing on those machines.

As a side note, more recently, a hybrid version of digital-analogue quantum computer \cite{parra-rodriguez_digital-analog_2020, martin_digital-analog_2020} emerged with promising results for QAOA \cite{headley_approximating_2022}. To this day, however, it was not widely adopted, and more recently, a package called Qadence \cite{seitz_qadence_2024} emerged. In general, both AQC and digital-analogue schemes remain at the experimental level without major software development kits. This presents a gap for future exploitation. Moreover, offerings from D-Wave focus only on a specific, restricted form of Ising Hamiltonian, and as such, their programming model only caters to this restricted model. However, in general, AQC is universal and equivalent to circuit-based QC. This hints that there is a need for programming models that allow for the practical exploitation of fully capable analogue quantum devices to realise AQC.

\subsection{Limitations}
\label{sec:limitations}
Our selection for specific use cases and applications limits the available hardware with desired capabilities. Because most of the existing hardware control software is tied to a specific platform, by extension, this also limits the corresponding control software. This is the natural limitation of this review. Furthermore, the analysis would benefit from the deeper benchmarking of the associated packages. One natural candidate for benchmarking would be the compilation performance. Nevertheless, due to the characteristics of most packages, this is currently impossible as the packages are either too low level and the resulting pulse sequence is directly submitted to the device, or the compilation details are hidden on the remote backend. This further highlights the need for open-source projects that handle instruction offloading to the device.

Another major limitation of this study is that the metrics or benchmarks presented here are qualitative and are based on the authors' judgment. Unfortunately, the space of quantum programming models is still at the experimental stage and is very dynamic. As such, there is lack of major code bases which use those technologies. The study would benefit from other metrics such as Source lines of code (SLOC), Cyclomatic Complexity \cite{mccabe_complexity_1976}, Halstead metrics \cite{halstead_elements_1977} and Fault Analysis of large-scale repositories. It is too early for that and deeper analysis must be done in the future.

However, we argue that this space is currently stuck in a cycle. For larger code bases to be produced, the integration challenges need to be solved first. Naturally, in the future, we expect that the range of tools will converge to a few major players, and more quantitative analysis should be possible. The study should be repeated in this space.

\section{Conclusion}
\label{sec:conclusion}
Quantum computing offers great promise of advantage on some classically hard problems. Nevertheless, to truly leverage this emerging technology, besides improvements in hardware, there is much work to be done in the software toolchain space. With the current low-level circuit programming model, it is clear that programming quantum computers at scale is challenging. An analogue quantum computing offers some remedy to this by possibly opening ways for higher-level abstraction constructs. The additional benefit of analogue techniques is the capability to leverage existing NISQ hardware better.

In this work, we have reviewed existing quantum software tools with analogue capabilities and examined frameworks that can run AHS and QA in detail. The conclusion is that the existing tools lack capabilities for seamless integration in HPC environments, making them hard to adopt for useful applications. 

While there is a large set of pulse-level hardware control libraries, using those for the general quantum community remains a niche application that requires extensive domain knowledge. As such, there is a need to develop new tools that scale and are easy to use, built for HPC and for scientists from other domains. 

Similarly, the research groups working on AHS made giant leaps into uncharted territories, probing regimes unattainable for classical computation. Nevertheless, we identify that the community should focus efforts on building programming models for integration with existing physical codes, such as codes used to simulate many-body systems, to speed up those and shift the technology from experimental setups towards production and industry.

\section*{Acknowledgements}
We thank the anonymous reviewers for their constructive feedback. Additionally, we would like to thank Dr. Joseph Lee for his helpful discussions during early stages of drafting this review.
The authors acknowledge support from the Engineering and Physical Sciences Research Council grant number EP/Z53318X/1.

\bibliographystyle{quantum}
\bibliography{qc-new}

\end{document}